\documentclass[a4paper,11pt]{article}
\pdfoutput=1 

\usepackage{jcappub} 

\usepackage[T1]{fontenc} 
\usepackage{comment}
\newcommand{\be}{\begin{equation}}
\newcommand{\ee}{\end{equation}}

\newcommand{\bea}{\begin{eqnarray}}
\newcommand{\eea}{\end{eqnarray}}
\numberwithin{equation}{section}
\newcommand{\todo}[1]{{\bf\emph{\small {#1}}}\marginpar{$\Longleftarrow$}}
\newcommand{\V}{{\cal V}}
\newcommand{\vp}{{\varphi}}
\newcommand{\tx}{{\tilde x}}
\newcommand{\tF}{{\tilde F}}
\newcommand{\tT}{{\tilde T}}
\newcommand{\ab}{{\tilde\alpha}}
\def\rmi{{\rm i}}
\title{\boldmath Finite temperature effects in modular cosmology}

\author{Diego Gallego}

\affiliation{Escuela de F\'isica, Universidad Pedag\'ogica y Tecnol\'ogica de Colombia UPTC,\\ Avenida Central de Norte, Tunja, Colombia.}

\emailAdd{diego.gallego@uptc.edu.co}

\abstract{We revisit the cosmological history in the presence of light moduli by including possible thermal effects in the scalar potential.
\\
The well known cosmological moduli problem regards initial energy stored in the moduli due to a misalignment from its final position during inflation. We show that finite temperature corrections to the scalar potential, in general, induce similar effects and these are likely to overcome the ones from the misalignment. This changes important parameters like the preferred window for the numbers of e-fold during inflation and the final reheating temperature in a model-dependent manner. The general implications are a longer late modulus dominated epoch and a larger final reheating temperature. 
\\	
We explore all the discussed elements in type-IIB superstring Large Volume Compactification with a K\"ahler inflationary scenario, where zero temperature results are known. An instability analysis, using a Floquet approach, is also performed for this explicit case finding strong indications of possible oscillon production around a nearly universal normalized critical temperature, where the Floquet exponents show a divergent behaviour. 
}
 
\begin{document} 
\maketitle
\flushbottom

\section{Introduction}

Superstring compactifications, and in general supersymmetric (SUSY) theories, should generally deal with tens and even hundreds of scalar fields usuali called moduli. For phenomenological reasons, given the constraints on fifth forces, moduli fields should be massive, implying a non trivial scalar potential $V$, and it should be such that at the critical point (minimum) describing our universe satisfies:
\begin{enumerate}
	\item A positive definite Hessian, i.e., no tachyons, ensuring the metastability of the vacuum \footnote{We will no consider possible quintessence scenarios}. 
	\item Hessian eigenvalues larger than ${\cal O}(30 TeV)$. Given their Planck suppressed interactions their decay rate $\Gamma\sim \frac{1}{8\pi}\frac{m_\phi^3}{M_P^2}$ induces a reheating temperature $T\sim\sqrt{m_\phi^3/M_P}$ which, to be consistent with Big Bang Nucleosintesis, should be larger than ${\cal O}(1 MeV)$ translating into the just mentioned condition \cite{Coughlan:1983ci,Coughlan:1984yk,Banks:1993en,deCarlos:1993wie}. This is usually known as the cosmological moduli problem.
	\item A positive nearly zero Vacuum Expectation Value (VEV). This value simulates the observed cosmological constant which from the observation of an accelerated cosmological expansion \cite{Perlmutter:1998np,Riess:1998cb}  is accepted to be positive though very tiny, $\Lambda =(2.846\pm 0.076)\time 10^{-122}M_P$ \cite{Aghanim:2018eyx}.\footnote{See however possible issues pointed out in \cite{Kang:2019azh} about assumptions taken in \cite{Perlmutter:1998np,Riess:1998cb}.} As shown below this is usually related to the requirement of supersymmetry (SUSY) to be spontaneously broken.
	\item Finite values for the moduli. Although non a generic statement we will deal with moduli characterizing either coupling constants or geometric information of the compact manifold where the strings live. Finite values warranty a geometric regime and finite couplings. The ubiquitous critical point at infinite values for the geometrical moduli imply a decompactification\footnote{By using using the word decompactification we have in mind geometric moduli, though the argument can be extended to other kind of moduli.} case, not of our interest.\footnote{We can be more precise about this statement by requiring a Kaluza-Klein scale larger than $10^{-30} M_P$ to be consistent with 4D Newtonian gravity tests down to lengths of millimetre order \cite{Kapner:2006si}. }
	\item Leave room to accommodate inflationary scenarios. Being string theory a candidate for a unified description of the universe, it should contain as well the best candidate to explain the observed Cosmic Microwave Background.
	\item Long lived under vacuum decay due to its metaestability. In general the presence of further vacua allows the decay to lower energetic ones but the mean life time should be at least as long as the observed life time of our universe. 
\end{enumerate}
During the last two decades a great deal of understanding of the so-called moduli stabilization programme have been achieved,  mostly in type IIB superstring compactifications in Calabi-Yau (CY) manifolds with the seminal ideas in \cite{Giddings:2001yu} and the specific KKLT realization in \cite{Kachru:2003aw}. The idea is to use an effective 4D approach, though in recent years there have been a lot of progress in understanding some of these mechanisms from a 10D perspective \cite{Moritz:2017xto,Hamada:2018qef,Gautason:2018gln,Gautason:2019jwq,Hamada:2019ack,Bena:2019mte,Kachru:2019dvo,Grana:2020hyu}, with an ${\cal N}=1$ Supergravity (SUGRA) theory for which the scalar potential, neglecting the D-term components easily included in our discussion, is given by
\begin{equation}\label{eq:FtermPot}
V_F=e^{K}\left(D_IW K^{I\bar J
}\overline{D}_{\bar J}\overline W-3|W|^2\right)\,,
\end{equation}
in $M_{Planck}=1$ units, with the covariant derivatives, $D_I W=\frac{\partial}{\partial \Phi^I}W+W\frac{\partial}{\partial \Phi^I}K$, and $K^{I\bar I}\equiv \left(\frac{\partial^2}{\partial \Phi^I\partial \overline{\Phi}^{\bar I}}K\right)^{-1}$ the inverse K\"ahler metric, the $\Phi^I$ running over all moduli. SUSY minima are dictated by $\langle D_I W \rangle=0$ and therefore have a negative cosmological constant (as far $\langle W\rangle \neq0$). 
\\
Obtaining a dS vacuum from string compactification has turned a great challenge and recently a series of ideas, commonly known as {\sl Swampland Conjectures}, rise doubts about their existence from consistent UV completions of gravity \cite{Obied:2018sgi}. However, besides some clear criticisms on the arguments  behind the conjectures and their consequences, c.f. \cite{Roupec:2018mbn,Conlon:2018eyr, Cicoli:2018kdo,Akrami:2018ylq,Marsh:2018kub,Murayama:2018lie,Choi:2018rze,Danielsson:2018qpa, Hebecker:2018vxz}, the level of maturity of the methods and understanding of the problems leads to take the conjecture with caution and not as a guidance. Indeed, following this 4D effective description several constructions has shown the possibility of finding such kind of vacua, c.f.
\cite{Ftermup0,Ftermup1,Ftermup2,Ftermup3,Ftermup4,Ftermup5,Ftermup6,Ftermup7,Ftermup8,Ftermup9,Ftermup10,Ftermup11,Ftermup12,Ftermup13,Dtermup1,Dtermup2,Dtermup3,Dtermup4,Dtermup5,Becker:2002nn,Kahlerup2,Kahlerup3,Kahlerup4,Conlon:2005jm,Cremades:2007ig,Krippendorf:2009zza,Cicoli:2012fh,Cicoli:2013mpa,deAlwis:2013gka,Cicoli:2015ylx,Gallego:2017dvd}. One of the main playgrounds for these constructions is the so-called Large Volume Scenarios (LVS) \cite{LVS1Conlon:2005ki,LVS2Cicoli:2007xp} where important hierarchies for phenomenology are explained from an exponential size of the compact manifold volume. 
\\
The moduli problem assumes that the decay of a light modulus eventually reheats the universe, implying initial an energy stored in the modulus. This is easily justified if there is a coupling between the inflaton and the modulus. Indeed, during inflation, the dynamics of the modulus are different from the ones after, when the inflaton relaxes to its minimum. Therefore the temporal critical point differs from the final one. Once inflation ends the initial critical point loses this property but still the modulus is pinned to this value due to Hubble friction, of order the inflationary scale. Once the Hubble parameter lowers down to the order the mass of the modulus, this last one starts to oscillate with amplitude $\delta\vp$ given by the misalignment between the two critical points. To this coherent oscillations an energy is associated of order $\rho_\vp\sim V'' \delta\vp^2$ where the derivatives are evaluated at the final critical point (for a recent discussion see \cite{Kane:2015jia,Acharya:2019pas}.)
\\
One further conclusion is that the cosmological history has an extra matter dominated epoch between the usual radiation dominated epoch and the final matter dominated era,\footnote{We restrict the discussion to the case of a single modulus with the afore mentioned properties. The analysis with several moduli follow the same lines.} since coherent scalar oscillation behave in this way. Each epoch is characterized by a number of e-folds describing the change in the scale factor in powers of the Euler number. For the inflationary era, the standard value lies in the range $50$ to $60$ e-folds, corresponding to the expansion between the moment of horizon exit, also known as pivot scale, and the end of inflation. However, this number depends on observables determining the horizon re-entry and the cosmological history after inflation. Therefore a precise statement on the number of e-folds for inflation needs a detailed study of the late evolution of the universe \cite{Liddle:2003as}.
\\
One aspect that is important to keep in mind is that the modulus evolves during an epoch on which a first reheating has taken place and possible finite temperature effects appear. In particular thermal corrections to the scalar potential change the profile on which the modulus is stabilized and oscillates. This might imply that the zero-temperature critical point completely disappears leading to a decompactification process. But even in the less dramatic case on which the minimum persists its position is changed. Then, in an analogous situation to the one with the misalignment presented above, an oscillation of the modulus is expected with an amplitude roughly of the size of the shift of the minimum due to the thermal corrections (see  \cite{Nakayama:2008ks} for similar ideas). One important fact on this observation is that it is independent of the initial misalignment and therefore it is generally present even when this last one is absent. Also, as discussed below, due to the redshift from the end of inflation until reheating the energy associated with the thermal shift likely turns out to be dominant compared to the one from the initial misalignment, stressing the importance of studying this effect. Thermal effects in moduli stabilization have been studied since the KKLT construction was proposed \cite{Buchmuller:2004xr,Buchmuller:2004tz} (see also \cite{Barreiro:2007hb,Anguelova:2007ex,Papineau:2008xf,Anguelova:2009ht}) and an explicit evaluation for the misalignment in LVS was recently done in \cite{Cicoli:2016olq}. The present work explores the complete picture, first in a general framework, and then in LVS within a K\"ahler inflation scenario \cite{Conlon:2005jm}.
\\
A possible general statement is that any effect in the cosmological history that depends on the scalar potential should be evaluated in the light of thermal corrections. One of such is the production of localized, long-lived, non-linear excitations of the scalar field, a.k.a. oscillons \cite{Copeland:1995fq}. These have been extensively studied in a general set up (c.f.  \cite{Amin:2010dc,Amin:2011hu,Amin:2013ika} and references therein) and in particular for string compactifications a detailed analysis was done in \cite{Antusch:2017flz}. This analysis strongly depends on the scalar potential and in \cite{Antusch:2017flz} thermal effects were neglected.
\\
The report has the following structure: the next section is devoted to reviewing some general aspects of moduli stabilization and characterization of vacua from string theory compactifications. In particular, we review how precise statements on the cosmological history depends on the details of the scalar potential; the third section introduces finite temperature corrections and discuses, in this general setup, how these might affect not only the moduli stabilization procedure but also de cosmological history, presenting in this way the main ideas in the work; section \ref{sec:LVSchap} illustrates all previous ideas implementing them in the explicit playground of LVS leaving the discussion as general as possible; next in section \ref{sec:Floquetan} we study possible instabilities from non-perturbative dynamics in presence of thermal corrections using Floquet approach; then we conclude with some discussion.

\section{Moduli stabilization and Modular Cosmology }\label{sec:generalmodstab}

Let us start with a general discussion where we simply regard the dynamics stabilizing the moduli as known. Among these dynamics, it turns useful to distinguish different components that we suppose are easy to be identified: for example, sometimes one can isolate the main source of positive energy that turns the cosmological constant nearly zero, usually called {\sl uplifting} term, i.e.,
\be
V=V_0+V_{uplift}\,.
\ee
Given the complexity of the scalar potentials one of the most used methods to find phenomenological interesting minima of the potential is to make the search only with the $V_0$ part, for example, SUSY solutions that are rather easy to spot. Then, by including the corrections from $V_{uplift}$, usually regarded as small, the wanted vacua are constructed. In general, one cannot warranty that the properties of both vacua, the SUSY, and the non-SUSY one, are similar and in any case, the minimization of the full scalar potential should be done. In the following, we suppose that we know the solution and properties of the desired vacuum, like the spectrum. This allows, for example, effective approaches like leaving in the game only the lightest modulus, that we denote by $\vp$, making in some cases clearer the analysis than the usual situation of tens or hundreds of fields appearing in this kind of compactifications.
\\
This one-dimensional situation, for example, allows discussing some usual details of the string vacua. The vacuum we are interested in, i.e., satisfying the conditions exposed in the introduction, is a critical point of the scalar potential that we denote by $\vp_*$.
\begin{figure}[htb]
	\begin{center}
		\includegraphics[width=0.6\textwidth
		]{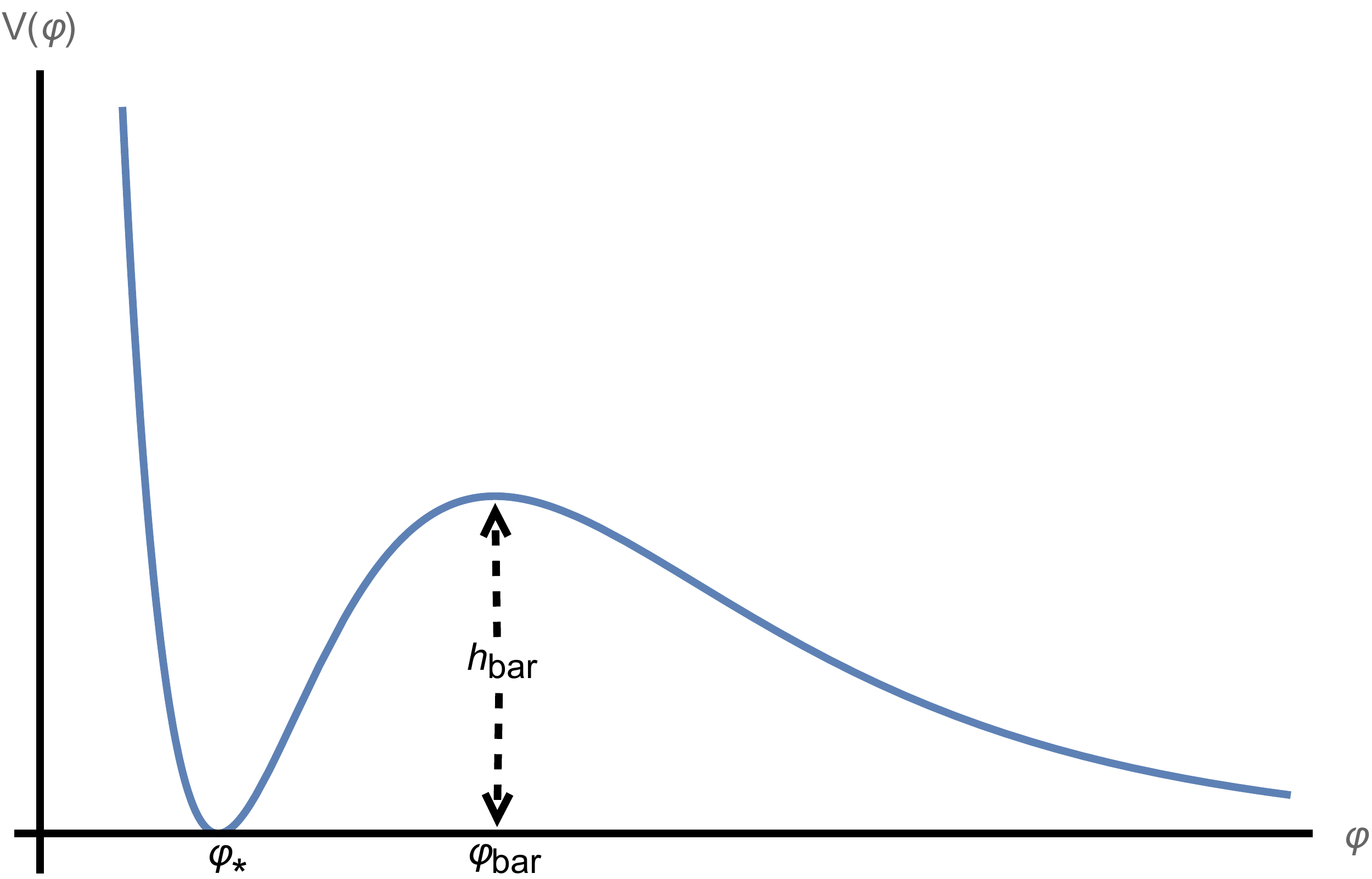}
	\end{center}
	\caption{Effective scalar potential with a single degree of freedom. The position of the minimum  $\vp_*$ and the barrier top $\vp_{bar}$ are shown as well the barrier height, $h_{bar}$.}\label{fig:genpot} 
\end{figure}
As argued in the introduction our universe is most likely a dS minimum of the potential and therefore is metastable, in particular, it can decay to the vacuum at $V(\vp\to \infty)\to 0$. Therefore there is a second critical point, that we denote $\vp_{bar}$, setting the place of a barrier potential separating our vacuum to the nearest minimum, that usually in simple models is just the decompactification one just mentioned, as illustrated in figure \ref{fig:genpot}.
\\
Roughly all these features are dictated by $V_0(\vp)$, responsible for the appearance of the critical point. However, the dynamics from the uplift contribution can be large enough for the rough estimations not to be reliable. In particular, the general structure for the uplift contribution is of the form\footnote{See \cite{Cicoli:2012fh,Gallego:2015vpa} for some explicit examples in the market.} $V_{up}\sim \vp^{-\gamma}$ which not only pushes the minimum towards the barrier but also decreases the convexity, i.e., the mass. Therefore, precise statements will need always a complete study of the scalar potential, at least at the effective level as the one we are suggesting.
\\
Inside $V$ there are dynamics generating an inflationary scenario which we regard effectively described by a single inflaton, $\tau$, accompanying the lighter modulus $\vp$. These two fields are expected to have couplings among them such that the dynamics of one affects the second; then for example during inflation, when the value taken by the inflaton is different to the actual one at the minimum, the value of $\vp$ is also different to $\vp_*$. A first estimation for such a misalignment can be done if the dynamics driving inflation can be isolated, i.e.,
\be
V(\vp, \tau)=V_{mod}(\vp)+V_{inf}(\vp,\tau)\,,
\ee
then, 
\be\label{eq:misaligmentgen}
\delta \vp\sim -\frac{V_{inf}'}{m_\vp^2}\,,
\ee
with the prime denoting a derivative respect to $\vp$ and $m_\vp^2\equiv \langle V_{0,mod}''\rangle $ the mass evaluated at the actual minimum and we regard that $V'_{inf}$ during inflation is negligible.
\\
This misalignment implies that once the Hubble parameter falls to values $H\sim m_\vp$ the modulus starts to oscillate with an amplitude $\delta\vp$. The energy associated with these coherent oscillations can be estimated to be
\be\label{eq:inmodenergyGen}
\rho_{mod}=\frac12 m_\vp^2\delta\vp^2\,,
\ee 
or more explicitly
\be\label{eq:rhomodend}
\rho^\vp_{end}\approx\frac12 \left(\frac{V_{inf}'}{m_\vp}\right)^2\,.
\ee
The subscript {\sl end} denotes that is the energy just after inflation ends, while the superscript indicates de corresponding energy component.
\\
Another possible consequence of this misalignment and further oscillations is an overshooting problem \cite{Brustein:1992nk}. Although probably less severe as naively expected, oscillations of the modulus might be such that it can overcome the barrier potential and reach the runaway region. A simple, but probably naive, way to avoid such a problem is to ask for the misalignment to be smaller than the distance from the minimum to the barrier top, i.e., with $\Delta\vp\equiv\vp_{bar}-\vp_*$ 
\be
\delta \vp <\Delta\vp \,,
\ee 
which might constrain the parameter space of a given model.

\subsection{Cosmological history in presence of moduli}\label{sec:cosmohistzeroT}

Cosmological history is divided into several epochs defined depending on the component of energy density that dominates. In the standard scenario the energy stored in the inflaton, first as a non-vanishing value for the scalar potential and then as coherent oscillations, is then transmitted via its decays to the rest of the matter reheating the universe, so far cooled down due to inflation. The reheating epoch lasts until thermalization takes place and we speak then after of a radiation dominated epoch when much of the particles are highly relativistic. This radiation energy density dilutes faster than the matter component and at some point, this last one starts to dominate. Since the cosmological constant does not dilutes soon or later it will dominate, as is the present case of our universe.
\\
As discussed already in the introduction the presence of light scalars might change this scenario with several matter-dominated epochs. Our analysis is restricted to only one of such fields being straightforward to extend to the general case.
\\ 
We closely follow \cite{Dutta:2014tya} with notation for the sub-indices {\sl end}, $\tau dec$, {\sl re}, {\sl eq} and {\sl dec} to denote the end of: inflation, inflaton-dominated, reheating, radiation-dominated (matter-radiation equality), and modulus-dominated epochs respectively. For each epoch $i$ we have a number of e-folds
\be\label{eq:efoldgen}
N_i=\ln\left(\frac{a_{i,end}}{a_{i,in}}\right)\,,
\ee
that can be related to the energy density using 
\be\label{eq:rhorelgen}
\rho(t)=\left(\frac{a_{i,in}}{a(t)}\right)^{3(1+\omega_i)}\rho_{i,in}\,,
\ee
with $\omega_i$ the equation of state parameter for the correspondent epoch. Then
\be
N_i=\frac{1}{3(1+\omega_i)}\ln\left(\frac{\rho_{i,in}}{\rho_{i,end}}\right)\,.
\ee

\subsubsection{Inflaton dominated epoch}
Just after the end of the inflationary epoch the energy density is dominated by inflaton quanta behaving like non-relativistic matter, i.e. $\omega_i=0$. Then 
\be
N_\tau=\frac13 \ln\left(\frac{\rho_{end}}{\rho_{\tau dec}}\right)\,.
\ee
With the assumption of an efficient inflaton quanta production the initial energy density can be associated to the one stored in the scalar potential, 
\be
\rho_{end}\approx V_{inf}=3 H^2_o M_P^2\,.
\ee
while the one at the moment of inflaton decay is given by the decay rate
\be
\rho_{\tau dec}\approx 3 M_P^2 \Gamma_\tau^2\,.
\ee
Therefore
\be\label{eq:taudomEfold}
N_\tau\approx\frac13 \ln\left(\frac{V_{inf}}{3M_P^2 \Gamma_\tau^2}\right)\,,
\ee
where we could be more explicit for example if consider that the inflaton interacts with gravitational strength such that $\Gamma_\tau\sim m_\tau^3/M_P^2$.

\subsubsection{Reheating epoch}

During this period the universe suffers a thermalisation process until a thermal bath is developed. This epoch can be characterised by a general equation of state with parameter $\omega_{re}$ where the standard scenario regards $\omega_{re}=0$, i.e., like matter, while under well-motivated argument is expected that $\omega_{re}<1/3$  and numerical studies usually scan in the range $-1/3\leq \omega_{re}\leq 1/3$ \cite{Mortonson:2010er,Bhattacharya:2017pws}. We leave it generic, thus we have for this epoch
\be\label{eq:Nre}
N_{re}=\frac{1}{3(1+\omega_{re})}\ln\left(\frac{\rho_{\tau dec}}{\rho_{re}}\right)\approx\frac{1}{3(1+\omega_{re})}\ln\left(\frac{3M_P^2 \Gamma_\tau^2}{\rho_{re}}\right)\,.
\ee
In much of the studies an instantaneous reheating, i.e., $N_{re}=0$, is regarded.

\subsubsection{Radiation dominated epoch}
Radiation domination is characterised by $\omega_{rad}=1/3$, therefore the scale factor can be written in terms of both the radiation energy density and the one corresponding to the modulus, as follows
\be\label{eq:aradamod}
\frac{a(t)}{a_{re}}=\frac{\rho^{rad}_{re}}{\rho^\vp_{re}}\frac{\rho^\vp(t)}{\rho^{rad}(t)}\,.
\ee
Radiation dominated epoch ends at equality, i.e., $\rho^{rad}_{eq}=\rho^{\vp}_{eq}$,  thus
\be
\frac{a_{eq}}{a_{re}}=\frac{\rho^{rad}_{re}}{\rho^\vp_{re}}\approx\frac{\rho_{re}}{\rho^\vp_{re}}\,,
\ee
and
\be\label{eq:Nradpre}
N_{rad}=\ln\left(\frac{\rho_{re}}{\rho^\vp_{re}}\right)\,.
\ee
We can further develop this expression using \eqref{eq:Nre}, such that
\be
\rho_{re}=e^{-3(1+\omega_{re})N_{re}}M_P^2\Gamma_\tau^2\,,
\ee
and also \eqref{eq:efoldgen} and \eqref{eq:rhorelgen} giving
\be
\rho^\vp_{re}=\left(\frac{a_{re}}{a_{\tau dec}}\frac{a_{\tau dec}}{a_{end}}\right)^3\rho_{end}=e^{-3(N_\tau +N_{re})}\rho^\vp_{end}\,.
\ee
The energy density associated to the modulus arises due to the misalignment and afterwards oscillation of the modulus around the minimum, with amplitude \eqref{eq:misaligmentgen}, was estimated to be \eqref{eq:inmodenergyGen} and \eqref{eq:rhomodend}.
\\
In here we neglect the delay between the end of inflation and the start of the modulus oscillation, happening just until $H\sim m_\vp$. Finally we get the implicit expression
\be\label{eq:Nrad}
N_{rad}=\ln\left(e^{3 N_\tau}M_P^2\Gamma_\tau^2\left(\frac{m_\vp}{\delta V'_\tau}\right)^2\right)-3 \omega_{re}N_{re}\,.
\ee
Another way of looking at $N_{rad}$ is to take $M_P^2\Gamma_\tau^2$ as $\rho^\tau_{\tau dec}=e^{-3N_\tau}\rho^\tau_{end}$, then the argument inside \eqref{eq:Nrad} is given in terms of the ratio between the energy stored in the inflaton and the one in the modulus
\be\label{def:epsilon}
\epsilon^2\equiv\frac{\rho^\vp_{end}}{\rho^\tau_{end}}\,.
\ee
With this consideration we get
\be\label{eq:Nrad2}
N_{rad}=-\ln\left(\epsilon^2\right)-3 \omega_{re}N_{re}\,.
\ee

\subsubsection{Modulus dominated epoch}

This epoch lasts until the modulus decay thus the analysis is similar to the one of the inflaton domination. In particular the energy at the end is given by
\be
\rho_{dec}\approx 3 M_P^2 \Gamma^2_\vp\,,
\ee
while, as discussed above,
\be
\rho_{eq}=e^{ -3(N_\tau+N_{re}+N_{rad})}\rho^\vp_{end}=3 M_P^2 \Gamma_\tau^2\epsilon^8e^{-(1-3\omega_{re})N_{re}}\,, 
\ee
where we used the results \eqref{eq:taudomEfold}, \eqref{eq:Nrad2} and the definition \eqref{def:epsilon} with $\rho_{end}^\tau=V_{inf}$. Then
\be\label{eq:NeModdomGen}
N_\vp=\frac43 \ln\left(\epsilon^2\right)+\frac23 \ln\left(\frac{\Gamma_\tau}{\Gamma_\vp}\right)-\frac13(1-3\omega_{re})N_{re}\,.
\ee
More explicitly regarding Planck suppressed interactions, i.e., $\Gamma\sim m^3/M_P^2$, we get
\be\label{eq:NeModdomGenSimp}
N_\vp=\frac43 \ln\left(\epsilon^2\right)+ \ln\left(\frac{m_\tau^2}{m_\vp^2}\right)-\frac13(1-3\omega_{re})N_{re}\,.
\ee
\subsection{Final reheating temperature}
To this aim, we can use the Hubble parameter at the moment the modulus decay. This is easily evaluated from the results so far and the general evolution
\be
H(t)=H_o\left(\frac{a_o}{a(t)}\right)^{-\frac32(1+\omega)}\,.
\ee
Then, with
\be\label{eq:Hubbleparameteratdec}
H_{dec}=\frac{H_{dec}}{H_{eq}}\cdot\frac{H_{eq}}{H_{re}}\cdot\frac{H_{re}}{H_{\tau dec}}\cdot\frac{H_{\tau dec}}{H_{end}}\cdot {H_{end}}\,,
\ee
we find
\be
H_{dec}=\Gamma_\vp e^{-(1-3\omega_{re})N_{re}}\,,
\ee
where we used $H_{end}=\sqrt{\frac{V_{inf}}{3M_P^2}}$. Then using the standard relation $3M_P^2H_{dec}^2=\rho_{dec}\approx \frac{\pi^2}{30}g_*T_{rh}^4$ we find\footnote{For the ease of notation we keep a single notation for the effective number of relativistic particles, $g_*$, being in general simply taken of order of hundreds and not dramatically changing the final results. But it is clear that this number should be evaluated at the required epoch.}
\be\label{eq:FinalTnThermal}
T_{rh}=\left(\frac{90}{\pi^2g_*}\right)^{1/4} \sqrt{M_P\Gamma_\vp} e^{-\frac14(1-3\omega_{re})N_{re}}\,,
\ee
with $g_*$ the effective number of relativistic species at the moment of reheating.
\\
Notice that we could as well consider some finite duration for the final reheating epoch, in which case we will have a generalised version for the one obtained in \cite{Bhattacharya:2017ysa}.

\subsection{Number of e-folds and cosmological observables}

Inflationary cosmology observables include the spectral index $n_s$ and the tensor to scalar ratio $r$. The evaluation of these in an inflationary scenario is done at the pivot scale, namely at horizon exit, and therefore depend on the number of e-folds inflation last. This last one, in turn, is linked to the post inflationary history through observables like the Hubble parameter at equality and the ratios of the Hubble radii \cite{Liddle:2003as}. From the previous discussion should be clear that a precise statement on the number of e-fold for an inflationary scenario depends on the presence and dynamics of the light moduli. Here we review the analysis done in \cite{Dutta:2014tya} expressing the results in a more convenient way for our interests.
\\
The commoving wave number $k$ mode from horizon can be written as
\be
k=a_k H_k=\frac{a_k}{a_{end}}\frac{a_{end}}{a_{re}}\frac{a_{re}}{a_{ra}}\frac{a_{ra}}{a_{eq}}\frac{a_{eq}}{a_{dec}} a_{dec} H_k\,.
\ee
Leaving this in terms of the number of e-folds we have the following relation 
\be\label{eq:NsAndPivot}
N_{mat}=N_k-N_\tau-N_{re}-N_{ra}-\ln(k)+\ln(a_{dec})+\ln(H_k)\,.
\ee
Similarly, starting from the energy density, we have
\bea\label{eq:Nmatrhos}
3N_{mat}=-3N_\tau-4N_{rad}-3(1+\omega_{re})N_{re}+ \ln\left(\frac{\rho_{end}}{\rho_{dec}}\right)\,.
\eea
Combining these expressions we have
\be
N_k +\frac14\left(N_\vp+N_\tau\right)+\frac14\left(1-3\omega_{re}\right)N_{re}=  \log \left(\frac{a_{dec}H_k}{k}\right) +\frac14\ln \left(\frac{\rho_{dec}}{\rho_{end}}\right)\,.
\ee
The right-hand side can be recast into observables: the Hubble parameter at the pivot scale can be either be understood as a function of the tensor to scalar fluctuation ratio, $r$, and the scalar fluctuation amplitude, $A_s$,
\be
H_k^2=\frac12\pi^2 r A_s M_P^2\,,
\ee
but also in terms of the energy density
\be
\rho_k=3 M_P^2 H_K\,;
\ee
On the other hand, as we saw in the previous section, the energy density at the time of modulus decay is related to the final reheating temperature, which in turn is linked to the temperature at the present epoch through the assumption of entropy conservation:
\be
T_{rh}\approx \left(\frac{13}{11 g_{s,*}}\right)^{1/3}\left(\frac{a_0}{a_{dec}}\right)T_0\,,
\ee
with $g_{s,*}$ the number of effective relativistic species at the moment of modulus decay. With this informations, using both approaches for $H_k$ we have, taking $a_0=1$,
\be
N_k +\frac14\left(N_\vp+N_\tau\right)+\frac14\left(1-3\omega_{re}\right)N_{re}=  \frac14\log \left(\frac{\pi ^4 r A_s   T_0^4g_*}{180 k^4 }\left(\frac{13}{11 g_{s,*}}\right)^{4/3}\right) +\frac14\ln \left(\frac{\rho_{k}}{\rho_{end}}\right)\,.
\ee
Using the actual data for the temperature $T_0\approx 2.73\,K\approx 1.92\times10^{-32} M_P$ \cite{Fixsen:2009ug}, $\ln(10^{10}A_s)\approx 3.04$ \cite{Aghanim:2018eyx,Akrami:2018odb} and the pivot scale chosen for the Planck data $k=0.05Mpc^{-1}\approx 2.62\times10^{-59} Mp$, we finally find
\be\label{eq:numEfolds}
N_k +\frac14\left(N_\vp+N_\tau\right)+\frac14\left(1-3\omega_{re}\right)N_{re}\approx 56.8+\frac14 \ln(r)     +\frac14\ln \left(\frac{\rho_{k}}{\rho_{end}}\right)\,.
\ee 
Some comments are in order:  the last logarithm compares the energy at the beginning and at the end of inflation which for very flat potentials, as the standard lore assumes, will lead to a negligible contribution. Therefore, the main correction of the l.h.s. value comes from the tensor to scalar ratio which from Planck data \cite{Akrami:2018odb} is at least $\ln(r)<-2$; this equation then leads to a preferred value for the number of e-folds $N_k$, which would depend on the number of e-folds during moduli dominance $N_{mod}=N_\tau+N_\vp$ and the one for the reheating period; the number $N_k$ is also related to observables, for example, the spectral index of scalar perturbations, $n_s$, thus the relation above might turn in the future into a tool to reduce the parameter space in models of inflation or even guideline for model building. For example, this kind of expressions has been used to extract preferred values for the equation of state parameter in explicit inflationary scenarios \cite{Bhattacharya:2017ysa}.
\\
In the light of our general results we have that the preferred value is roughly given by
\be\label{eq:numEfoldsZeroT}
N_k\approx57-\frac{1}{12} \log \left(\frac{V_{inf} \epsilon ^8}{3 \Gamma _\tau ^2 M_P^2}\right)-\frac14\ln \left(\frac{m_\tau ^2}{m_\vp^2}\right)-\frac16(1-3\omega_{re})N_{re}+\frac14 \ln(r)     +\frac14\ln \left(\frac{\rho_{k}}{\rho_{end}}\right)\,,
\ee
which most likely will turn into a lower value to the standard used range of $50\leq N_k\leq 60$ \cite{Cicoli:2016olq}. Indeed, this relation, in a slightly different form, was used in \cite{Das:2015uwa} to get constraints on the moduli masses.

\section{Thermal corrections and modular cosmology}\label{sec:ThermalSec}

Even if light moduli are not directly coupled with the inflaton the decay of this last one might affect the dynamics of the first one via reheating and finite temperature effects. Indeed, after inflaton quanta domination its decay into light particles generates a thermal bath which, in turn, induces thermal corrections to the scalar potential. For the moment, for us it will be irrelevant if the thermal bath comes from a standard reheating scenario or a more sophisticated like preheating \cite{Kofman:1997yn} and even diverse as warm inflation \cite{Berera:1995ie,Berera:1995wh}. Later on we will take a closer look now at a specific scenario of inflation and using some considerations about the reheating process.
\\
Thermal corrections to the scalar potential come from finite temperature effects in the effective action. At one loop, and for high temperatures compared with the masses involved, this takes the form \cite{Kapusta:2006pm} (see \cite{Binetruy:1984wy,Binetruy:1984yx} for details with SUGRA theories),
\be\label{eq:V1loopSUSY}
V_1^T=-\frac{\pi^2 T^4}{90}\left(g_B+\frac78 g_f\right)+\frac{T^2}{24}\left( TrM_b^2+TrM_f^2\right)+{\cal O}(T M^3)\,,
\ee
where the subscripts $B$ and $f$ hold for bosons and fermions and the $g$'s in the parenthesis refer to the number of degrees of freedom in the thermal bath. The mass matrices in the $T^2 $ term correspond to the fields in the thermal bath and depend on the moduli. The 2-loop potential, showing deviations from the ideal gas approximation, takes the form
\be\label{eq:V2loopSUSY}
V_2^T=a T^4\left(\sum_j f_j(g_j)\right)+b T^2 \left(Tr M_b^2+TrM_f^2\right)\left(\sum_j f_j(g_j)\right)\,,
\ee
with the parameters $a$ and $b$ model dependent ${\cal O}(1)$ numbers. The functions in the sums are also model dependent whose arguments are the couplings involved in the respective Feynman diagram. Being the coupling moduli dependent these fields also appear in the $T^4$ term.
\\
Moduli dependency comes from the mass matrices and the couplings in the 2-loop $T^4$ terms. However, since the thermal bath is formed only by fields lighter than the temperature the leading dependency comes from the latter one. This as far as the relevant moduli indeed appear on it and that the couplings in the diagrams generating the second term in the 2-loop contribution are not large enough to compete with the scaling of the temperature, something possibly discarded from the perturbative approach from which this result comes from. 
\\
In order to make precise statements, and following \cite{Buchmuller:2004tz} (see also \cite{Anguelova:2009ht}) we regard $f(g)\propto g^2$ and that the gauge coupling in the thermal bath is dictated by the non-axionic part of the modulus, i.e., $g^2\sim1/\vp$. Then, the leading finite temperature contribution to the scalar potential takes the form
\be\label{eq:thermalscalarpotential}
V_T=T^4\left (\frac{\kappa}{x}-\kappa_1\right)\,,
\ee
with positive parameters of ${\cal O}(10)$.

\subsection{Moduli stabilisation}\label{sec:modulusthermaleff}

The previous observation leads us to consider corrections to the scalar potential that are schematically included as follows,
\be
V_{full}=V+V_T(\vp)\,.
\ee
The first thing to notice is that thermal corrections have a runaway profile and therefore its analysis follows closely the one done for the uplifting terms. In particular, it might lead to a decompactification if the temperature is too large.
\\
To estimate this maximal temperature we look for temperatures for which the dynamics from the $V_T$ around the minimum overcome the ones from the zero temperature scalar potential, although studies show that this can be a too restrictive point of view \cite{Barreiro:2007hb}. This can be done in several ways: the most direct is simply to compare $V_T$ with the barrier height separating the vacuum with the runaway region. Then denoting this by $h_{bar}$ we have
\be\label{eq:Tmaxbargen}
T_{max,bar}^4\approx\left |\frac{\kappa}{\vp_*}-\kappa_1\right|^{-1}h_{bar}\,,
\ee
where we used the explicit expression \eqref{eq:thermalscalarpotential} and as a first approximation we evaluate the thermal corrections at the original minimum; A second possibility is to pinpoint approximately the temperature at which the modulus mass turns zero, evidence of a saddle point, and therefore
\be\label{eq:Tmaxmassgen}
T^4_{max,mass}\approx \frac{\vp_*^3}{2\kappa}m_\vp^2\,,
\ee 
where the mass appearing in the expression is defined above as the one at zero temperature; a third approach is related to the main point in our study: thermal corrections change the position of the minimum shifting it towards the barrier. Then, the decompactification temperature can be defined as such that such a shift coincides with the distance from the minimum to the barrier.
\\
Following the same ideas used for the misalignment, we have that the shift due to thermal corrections can be estimated to be
\be\label{eq:thermalshift}
\delta_T\vp\approx -\frac{V_T'}{m_\vp^2}=
\frac{\kappa T^4}{\vp_*^2m_\vp^2}\,.
\ee
Then if the distance to the barrier is given by $\Delta\vp$ we have a further expression for the decompactification temperature
\be\label{eq:Tmaxshiftgen}
T_{max,shift}^4\approx \frac{\vp_*^2 \Delta \vp }{\kappa}m_\vp^2\,.
\ee
Notice that the first approach is equivalent to spot a temperature for which the barrier top sits at Minkowski, while the transient minimum is an AdS, and given the dominance of the negative contribution from thermal corrections (see section \ref{sec:radenergy} bellow) this situation is likely to happen before the other two occur. Therefore the first estimation would be the most conservative one.

\subsection{Cosmological history}

In light of the new dynamics from thermal corrections, we should revisit the analysis in section \ref{sec:cosmohistzeroT} after the reheating epoch. 
\\
In particular, the result in \eqref{eq:Nrad2} neglects that once the thermal bath settles down the minimum is shifted by \eqref{eq:thermalshift} and a kick on the modulus oscillation is expected. This results in an increase on the energy stored in the modulus. Indeed, we need to compare the energy associated to this new misalignment 
\be\label{eq:rhophireT}
\rho^\vp_{re,T}=\frac12 m_\vp^2 \delta_T\vp^2\sim \frac12 \frac{\kappa^2 T^8}{\vp^4 m_\vp^2}\,,
\ee
with the original one at the end of the reheating epoch. To proceed with this comparison we will make the approximation that the actual temperature appearing in $\delta_T\vp\sim \delta V_T'/m_\vp^2$ is precisely the maximal one. In any case, the energy densities to be compared depend on the initial misalignment and the thermal shifts, both of which are of Planck order. Therefore, the ratio is roughly given by the redshift suffered by the initial energy density, i.e.,
\be\label{def:theta}
\theta^2 \equiv\frac{\rho^\vp_{re}}{\rho^\vp_{re,T}}=\left(\frac{\delta\vp}{\delta_T\vp}\right)^2e^{-3(N_\tau+N_{re})}\sim \frac{V_{inf}}{3M_P^2 \Gamma_\tau^2} e^{-3N_{re}}\,.
\ee
\subsubsection{Radiation dominated epoch}
The outshot is that the radiation domination epoch is shortened while the late modulus dominated one gets larger. For the number of e-fold for the radiation dominated epoch we have now, in contrast to the result in \eqref{eq:Nrad2}, with the definition \eqref{def:epsilon}
\be\label{eq:NradT}
N_{T,rad}=\ln\left(\frac{e^{-3 \omega_{re}N_{re}}\theta^2}{ \epsilon^2}\right)\sim- \ln\left(\epsilon^2\frac{V_{inf}}{3M_P^2 \Gamma_\tau^2}\right)-3(1+ \omega_{re})N_{re}\,,
\ee
namely
\be
N_{T,rad}\sim N_{T=0,rad}+\ln\left(\theta^2\right)<N_{T=0,rad}\,,
\ee
since $\theta^2<1$. Naively it seems that easily the numbers of e-folds can be zero, i.e. no radiation domination epoch, meaning $\rho^{rad}_{re}\leq \rho^\vp_{re}$. However, with $\rho_{rad}=\frac{\pi^2}{30}g_*T^4$, this implies
\be
\frac{\pi^2g_*}{15\kappa^2}\vp^4 m_\vp^2\leq T^4 \,.
\ee
From the analysis in section \ref{sec:modulusthermaleff} we can relate the modulus mass to the maximal temperature, $T^2_{max}\sim \frac{\vp^3}{2\kappa}m_\vp^2$, thus in order the modulus component to be larger than the one from radiation from the beginning it should happen
\be
\frac{2\pi^2}{15}\frac{g_*}{\kappa}\vp_*T_{max}^4\leq T^4\,,
\ee
something that look quite difficult to achieve given that the parameter accompanying $T_{max}^4$ is expected to be of order  $g_*={\cal O}(100)$. The conclusion is that in our rough final estimation we are missing details that might be as important as the one considered. Still, the general conclusion about a shorter radiation dominated epoch holds since is a natural consequence of the increase of the modulus energy density.

\subsubsection{Modulus dominated epoch and thermal effects}
In case thermal effects are relevant we have
\be
\rho_{T,eq}=e^{ -3N_{T, rad}}\rho^\vp_{re,T}\,,
\ee
leading to the following number of e-folds,
\be\label{eq:Nphisthermal}
N_{T,\vp}=\frac13 \ln\left(\frac{\rho_{re,T}^\vp}{3M_P^2\Gamma_\vp^2}\right)-N_{T,rad}\,,
\ee
or
\be\label{eq:NmodT}
N_{T,\vp}\approx\frac13 \ln\left(\frac{\kappa^2 M_P^2T^8}{6 \vp_*^4 m_\vp^8}\right)-N_{T,rad}\,.
\ee
In general, using again the parameter $\theta$ in \eqref{def:theta} we have that
\be\label{eq:Nphisthermalrelation}
N_{T,\vp}=N_{T=0,\vp}-\frac13\ln\left(\theta^2\right)>N_{T=0,\vp}\,.
\ee
We stress the fact that this last relation is only in case both are present, for the modular domination is expected, lasting a number of e-folds given by \eqref{eq:NmodT}, even in the absence of an initial misalignment.

\subsubsection{Final reheating temperature}
With our results regarding the finite temperature corrections we have now that
\be
H_{dec,T}=H_{dec,T=0}\left(\theta^2\right)^{-3/2}\,,
\ee
leading to a reheating temperature
\be\label{eq:FinalTwThermal}
T_{rh,T}=T_{rh,T=0}/\theta^{3/4}>T_{rh,T=0}\,,
\ee
compared to the one obtained in \eqref{eq:FinalTnThermal}. This increase in the final reheating temperature might help in ameliorating possible tensions between this and $T_{BBN}$.

\subsubsection{A comment on the radiation energy density} \label{sec:radenergy}

Finite temperature corrections to the scalar potential are computed evaluating the quantum effective action \cite{Dolan:1973qd}, which is interpreted as a Free energy for the theory and can be read as the radiation pressure of the thermal bath. A proper conceptual treatment of this extra component implies to disentangle this term with the stress-energy tensor from the modulus. Indeed, the associated energy density comes from the radiation of the thermal bath, related to its pressure through
$$
\frac{d p^{rad}}{dT}=\frac{\rho^{rad}+p^{rad}}{T}\,.
$$
Therefore, in our case, where we regard terms beyond the ideal gas approximation we have
\be
\rho^{rad}=-3\left (\frac{\kappa}{\vp}-\kappa_1\right)T^4=\frac{\pi^2}{30}g_*T^4\left(1-\frac{90\kappa}{\pi^2 g_*\vp} \cdots\right)\,,
\ee
where $g_*^{re}=g_B+\frac78g_f$ and we identify in the first term the usual radiation energy density while the ellipses contain further moduli dependent terms coming from the neglected terms in the scalar potential.\footnote{Reference \cite{Barreiro:2007hb} developed on the same point but they unnecessarily distinguish two kinds of particles in the thermal bath. } Notice that this relationship, in particular, implies that for the energy density to be positive definite the negative contribution to the scalar potential from thermal corrections should be always larger than the moduli dependent one.

\subsection{Inflationary number of e-folds an thermal corrections}

\begin{figure}[htb]
	\begin{center}
		\includegraphics[width=0.6\textwidth
		]{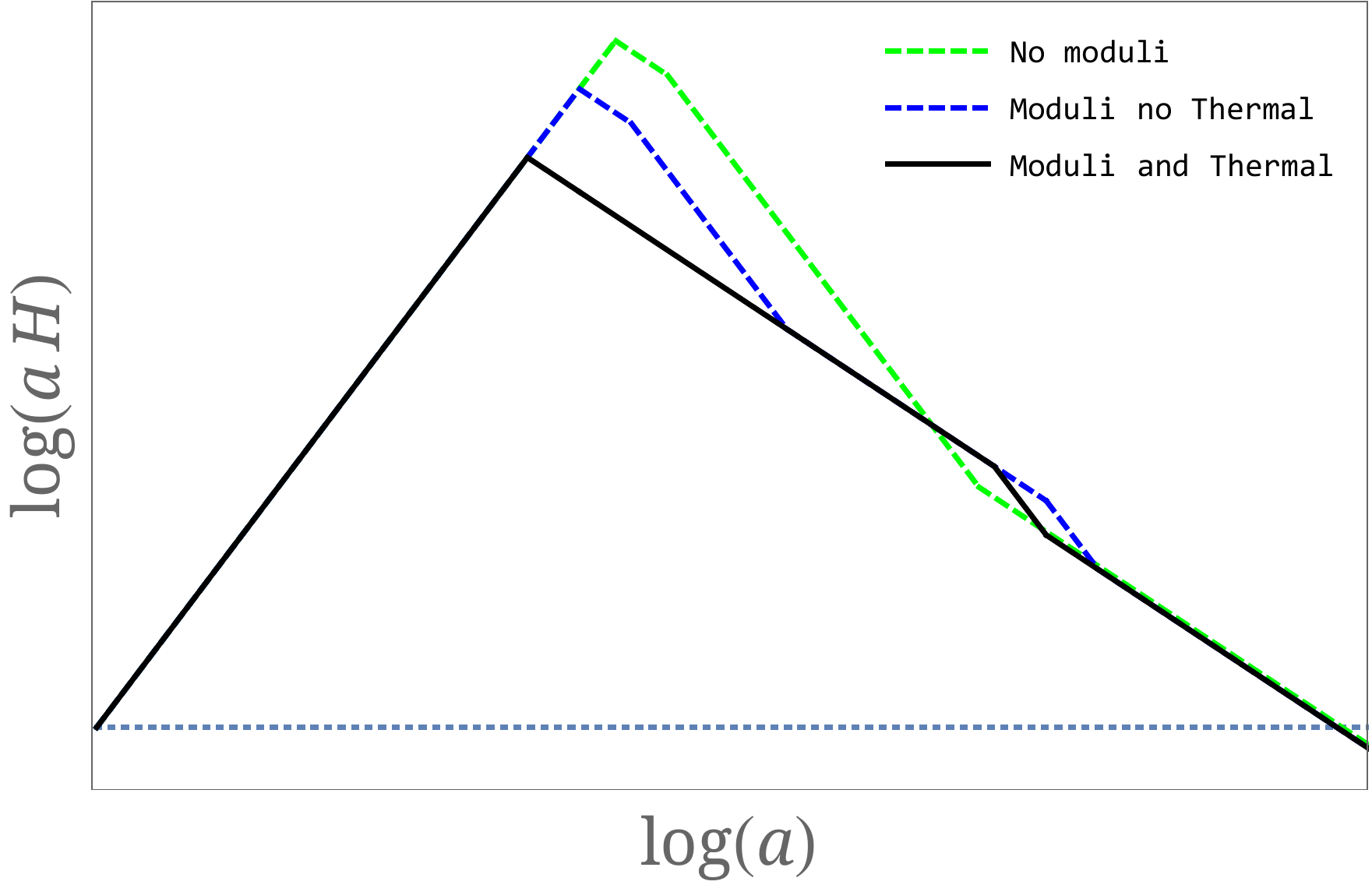}
	\end{center}
	\caption{Evolution of the Hubble radius for three cases: without light moduli, with light moduli with an initial misalignment and light moduli with thermal corrections. For the reheating phase, it is taken $\omega_{re}=0$ such that the slope during this epoch coincides with the one of matter domination. For the case of moduli with finite temperature effects, we show the extremal case with no radiation dominated period after the first reheating. A proper horizon re-entry implies a shorter period of inflation. }\label{fig:hrad} 
\end{figure}

The findings above also change the expression found for the number of e-folds during inflation \eqref{eq:numEfolds}. From \eqref{eq:Nphisthermal} we find
\be\label{eq:numEfoldsT}
N_k\approx57-\frac{1}{12} \log \left(\frac{\kappa ^2 T^8 V_{inf}}{18 \Gamma_\tau ^2 m_\vp^8 \vp_* ^4}\right)-\frac14(1-6\omega_{re})N_{re}+\frac14 \ln(r)     +\frac14\ln \left(\frac{\rho_{k}}{\rho_{end}}\right)\,,
\ee 
or with the relation \eqref{eq:Nphisthermalrelation}
\be
N_{k,T}=N_{k,T=0}+\frac{1}{12}\ln\left(\theta^2\right)<N_{k,T=0}\,,
\ee
showing, as expected, that thermal corrections will induce even smaller values for the preferred number of e-folds, compared to the ones just expected from misalignment. This is illustrated in figure \ref{fig:hrad}.

\section{General Large Volume Scenario}\label{sec:LVSchap}

The scenario originally proposed in \cite{LVS1Conlon:2005ki,LVS2Cicoli:2007xp} and coined as Large Volume Scenario combine three main ingredients from CY orientifold compactifications for type-IIB superstrings, namely: non-perturbative corrections to the effective superpotential, quantum corrections to the K\"ahler potential and multi moduli dynamics. The main aspect of such string vacua is that one of the moduli, characterizing the overall CY volume, is hierarchically larger than the other ones, but still regarded in the geometric regime. The simplest models of this kind of scenario can be analytically constructed regarding a Swizz-Chess like CY \footnote{Generalizations can be however worked out with fibrered CY \cite{Cicoli:2008gp}.} for which the volume takes the form
\be
\V\propto T^{3/2}-\sum_i\kappa_i t_i^{3/2}\,,
\ee
with $T$ and $t_i$ the non-axionic components of the K\"ahler moduli, the $t_i$ playing the role of the blow-up cycles visualized as the small holes of the Cheese. The K\"ahler potential reveals a non-scale structure with the following dependence on the moduli
\be\label{eq:Kalerpotgen}
K_{mod}={\cal K}-2\log\left(\V+\zeta\right)\,.
\ee
The parameters ${\cal K}$ and $\zeta$ are actually dilaton and complex structure moduli dependent which, however, are taken as stabilized via perturbative fluxes at supersymmetry preserving points such that their dynamics are irrelevant for the low energy studies usually regarded for phenomenology \cite{Gallego:2011jm,Gallego:2013qe} (see however \cite{Gallego:2017dvd}). The $\zeta$ term is actually an alpha prime correction \cite{Becker:2002nn} that will play a central role in the constructions of such vacua.
\\
The super potential has two main contributions, the classical perturbative part from fluxes \cite{Giddings:2001yu} and non-perturbatives components,
\be\label{eq:Wgen}
W_{mod}=W_{flux}+W_{np}(t_i)\,.
\ee
The first part for type-IIB compactifications only depends on the dilaton and complex structure moduli and actually is responsible of their stabilization. With the same argument above these moduli are regarded as fixed at a constant value, then $W_{flux}$ is just as a complex number and the same is true for possible moduli dependent parameters in $W_{np}$\footnote{See a detailed analysis of such a consideration in \cite{Gallego:2011jm,Gallego:2013qe}.}. Notice that only the small moduli $t_i$ enter in the superpotential. This due to the considerations that the non-pertubative contributions are exponentially suppressed with the schematic form $e^{-a t}$ making completely irrelevant the contributions from the large modulus.
\\
These ingredients are then used in the scalar potential \eqref{eq:FtermPot} and its details depend on the specification of each case but in general in an expansion in powers of the volume. We have three kinds of terms:
\begin{enumerate}
	\item A single term proportional to the alpha prime corrections and the modulus of the superpotential suppressed by the third power of the volume, i.e., $\sim \zeta |W|^2/\V^3$.
	\item Positive definite contributions from the terms $\partial_IW_{np} K^{I\bar J
	}\partial_{\bar J}\overline W_{np}$ mixing two non-perturbative contributions and suppressed by a positive power of  the volume.
	\item  Negative definite contribution from the terms $\partial_IW_{np} K^{I\bar J
	}\overline W_{flux}\partial_{\bar J}K+h.c.$ mixing one non-perturbative contribution with a perturbative one. These terms are as well suppressed by a positive, but different, power of the volume.
\end{enumerate}
The mentioned signs regard the axionic components already aligned such to get a minimum. For the first-mentioned term a requirement for this kind of vacua is that it turns to be positive, i.e., $\xi>0$, a condition that is then translated to possible topologies of the CY\footnote{This parameter is proportional to $\xi = -\frac{\chi(CY) \zeta(3)}{2 (2\pi)^3}$ with $\chi(CY)$ the Euler characteristic of the compact manifold.}.
\\
The idea is then the following: the interplay between the terms depending on the small moduli $t_i$, i.e., the ones coming from non-perturbative dynamics, stabilize them. This requires that both kinds of terms be of the same order of magnitude and implies a relation between the size of the volume and the exponentials; the term coming from the $\alpha$-prime corrections then is regarded as positive and renders the potential with a minimum in the volume modulus direction. This minimum given the leading non-scale structure of the scalar potential is non-supersymmetric but however in general fails in being a de Sitter one.
\\
This last observation implies that further dynamics should be included for this kind of vacua to turn phenomenologically attractive. Fortunately, as said in the introduction, these are easily found in string compactifications and, although their implementation is not straightforward, several explicit instances have been build-up for this scenario (cf.\cite{Conlon:2005jm,LVSMatter,Krippendorf:2009zza,Cicoli:2012fh,Cicoli:2013mpa,deAlwis:2013gka,Cicoli:2015ylx,Gallego:2017dvd}.)
\\
Moreover, thanks to the role played by the powers of the volume in labeling different terms in the scalar potential in LVS it is easy to spot this uplifting contribution just in the way they were discussed in section \ref{sec:generalmodstab} as studied in \cite{Gallego:2015vpa}.

\subsection{Features of nearly Minkowski dS LVS vacua} \label{sect:charLVS}

In the following, we consider the two K\"ahler moduli case, denoted by $x$ and $\V$ disregarding the axionic components being always possible to choose the parameters such that these have a null VEV, where the dynamics for all other possible fields are encoded in the numerical parameters, taken as positive, in the following scalar potential
\be\label{eq:LVSpot}
V=\frac{F}{ \V^{\gamma }}+\frac{\zeta }{ {\cal V}^3}+\frac{\lambda  e^{-2 x}  x^{\beta }}{ \V^{2 \eta-3 }}-\frac{\mu  e^{-x}  x^{\alpha }}{ \V^{\eta }}\,.
\ee
The powers in the exponentials are such to match their origin as explained before, i.e., linear and quadratic in the non-perturbative effects and absorbing possible factors in the definition of the modulus $x$. The corresponding powers for the volume are such that after the minimization and balance of both non-perturbative terms, leading to $e^x\sim \V^{3-\eta}$, these two scale like $\V^{-3}$. This potential is meant to be valid only for $x\gg1$, where multi-instantonic contributions can be neglected, and $\V$ of exponential size, such that possible further terms can be consistently neglected.\footnote{See appendix \ref{app:explicitUVLVS} for two explicit scenarios were we identify these effective parameters with microscopic ones from the K\"ahler potential and the superpotential.}
\\
It is clear that $\partial_{xx}V\gg \partial_{\V\V}V$ at any point in the moduli space where $\V\gg x$, which means in particular that the mass scale for the $x$ modulus larger than the one related to the volume modulus, a fact that is not going to change after canonical normalization. We can, therefore, follow an effective analysis where the only degree of freedom that is kept is the volume one.\footnote{The following analysis can be done as well in the UV description with the very same conclusions \cite{Gallego:2017dvd}.}
\\
Using the equation of motion (e.o.m.) for the small modulus we have
\be
e^{-x}= \frac{\mu  \V^{\eta -3} (x-\alpha ) x^{\alpha -\beta }}{\lambda  (2 x-\beta )}\approx  \frac{\mu  \V^{\eta -3}  x^{\alpha -\beta }}{2\lambda  }\,,
\ee
after regarding $\alpha$ and $\beta$ to be ${\cal O}(1)$. Then at leading order in $\ln\V$, taking $3-\eta>0$ we have the effective scalar potential
\be
V\approx \frac{F}{\V^{\gamma }}+\frac{\mu ^2 \left(x_0^{2 \alpha -\beta }-\ln ^{2 \alpha -\beta }\left(\V^{3-\eta }\right)\right)}{4 \lambda  \V^3}\,,
\ee
where we introduced $x_0= \left(\frac{4\zeta  \lambda }{\mu ^2}\right)^{\frac{1}{2 \alpha -\beta }}$, the leading analytic solution for $x$.
\\
Being interested for the moment in possible relations between the terms we rescale the amplitude $F=\frac{\tilde F \mu ^2 (3-\eta)^{2\alpha-\beta}}{4 \lambda }$ to get the following generic relative potential
\be 
V_{rel}=\frac{\tilde F }{ \V^{\gamma }}+\frac{\tilde x_0^{2 \alpha -\beta }-\log ^{2 \alpha -\beta }\left(\V\right)}{\V^3}\,,
\ee
The final outshot is that the parameters $\mu$, $\lambda$ $\eta$ and $\zeta$ only appear implicitly through  $\tilde x_0=x_0/(3-\eta)$. This however simply sets some units for the volume, as can be seen directly by writing $\tilde x_0=\ln\left(\V_0\right)$ such that any relation to size is confined in the constant $\V_0$. Notice moreover that only the combination $2\alpha-\beta$, that we regard positive definite, appears and we denote it hereafter by $\ab\equiv 2\alpha-\beta$.
\\
From this potential the e.o.m. reads
\be
-\gamma  \frac{\tF}{\V^{\gamma +1}}-\frac{\ab \log ^{\ab -1}(\V)}{\V^4}+\frac{3 \log ^{\ab }(\V)}{\V^4}-\frac{3 \tx^\ab}{\V^4}=0\,,
\ee
which implies the following cosmological constant and mass squared (non canonical),
\bea
\langle V_{rel}\rangle&=&\frac{1}{3} \left(-\frac{\ab \log ^{\ab -1}(\V_*)}{\V^3}+(3-\gamma) \frac{\tilde F}{ \V^{\gamma }_*}\right)\,,\cr
m_\V^2&=&\frac{3\ab  \log ^{\ab -1}(\V_*)-(3-\gamma) \gamma  \tilde F \V^{3-\gamma}_*}{\V^5_*}\,.
\eea
Then, in absence of uplift, $F=0$, the cosmological constant is negative but for $\gamma<3$ \footnote{The case $\gamma=3$ is excluded from the analysis since it can be absorbed in the original $\V^{-3}$ term while larger values most probably imply a different type of vacua where the $\V^{-3}$ and $\V^{-\gamma}$ terms will flip roles.} this term can make it zero or positive if we choose $\tilde F\sim \V^{\gamma-3}$, i.e., the fine-tuning would be less dramatic for large values of $\gamma$. On the other hand, the contribution from the uplift to the mass is negative, as is expected for of a runaway type potential. Therefore the uplifting amplitude cannot be extremely large or otherwise destabilizes the potential. Another way to say this is that the two solutions that appear for the e.o.m., i.e., a minimum and a maximum at the barrier top, degenerate for this extreme value of $\tilde F$. Then, requiring at least a Minkowski vacuum and a positive mass squared, we have a working window for the uplift amplitude:
\be
1\leq \frac{(3-\gamma) \V_*^{3-\gamma } \log ^{1-\ab}(\V_*)}{\ab } \tilde F< \frac{3}{\gamma}\,.
\ee
This tells us that for $\gamma<3$ there is always a possible value for the amplitude $F$ to have a metastable nearly Minkowski local minimum. Notice that this statement is completely general so, at this level of approximation, leading in $\ln\V$, the microscopic parameters are irrelevant. This conclusion breaks down when any parameter starts to be as large as $x\sim \ln\V$  in which case the window might get narrower.
\\
Being our universe nearly Minkowski let us take the lower bound and study the properties of the vacuum compared to the original AdS one. It is convenient to work from now on with the canonically normalized field $\phi=\sqrt{\frac32}\ln\V$ but for ease of notation we just write everything in terms of $\vp=\ln\V$
\be
V_{rel}=\tF e^{-\gamma  \varphi }+e^{-3 \varphi } \left(\tx^\ab_0-\varphi ^\ab\right)\,,
\ee
for which the e.o.m. reads,
\be
e^{-3 \varphi } \left(\varphi ^{\ab -1} (3 \varphi -\ab)-3 \tx^\ab_0\right)-\gamma  \tF e^{-\gamma  \varphi }=0\,.
\ee
The original solution, $\vp_0$, previous the uplift, i.e., $\tilde F=0$, is found in an expansion in $\tilde x\gg1$,
\be
\vp_0\approx \tx +\frac13\,.
\ee
Once the $\tilde F$ value for Minkowski is used the solution instead is
\be
\vp_*\approx\tx+\frac{1}{(3-\gamma)}\,,
\ee
where it is clear that at first approximation, with $\tilde x\gg1$, it is close to the non uplifted vacuum. This shift, $\vp_*-\vp_0=\frac{\gamma }{3 (3-\gamma)}$, implies in the overall volume a  relation
\be
\V_*\approx e^{\frac{\gamma }{3 (3-\gamma)}}\V_0\,,
\ee
or a relative change
\be
\frac{\Delta\V}{\V}\approx e^{\frac{\gamma }{3(3- \gamma)} }-1\,.
\ee
The exponential dependency on the value of $\gamma$ can dramatically change the scale energies implied in the solution, which in principle where only controlled by the value of $\tx$ trough $e^{\tx}$. Notice, for example, that for a value $\gamma=14/5$, proposed in \cite{Cicoli:2013mpa,deAlwis:2013gka}, the change on the volume is of two orders of magnitude, that for well-motivated values of the volume $\V\sim 10^5-10^7$ \cite{Conlon:2005jm,Anguelova:2009ht} might imply important changes in the parameter space compared with the naive expectations.
\\
On the phenomenological side, however, are the ratios rather the values itself what matter so no relevant changes seem to appear. To be precise the value of the volume at the actual vacuum controls the scale of all masses but, as we already notice, the light mode gets a further reduction from the parameter $\gamma$. Indeed, the flattening of the potential due to the runway uplift contribution makes the scalar potential convexity smaller. In general, we have the following mass for the volume modulus at the Minkowski vacuum\footnote{The physical mass still misses a factor $\frac23$ from canonical normalization and the overall term $\frac{\mu ^2 (3-\eta)^{2\alpha-\beta}}{4 \lambda }$.}
\be
m_\vp^2=\frac{ (3-\gamma)\ab  \tx^{\ab -1}}{\V_*^3}\,,
\ee
with $\V_*=e^{\vp_*}$. Since the case $\gamma=0$ does not change the profile of the potential it coincides with the mass in the original AdS solution. The term $(3-\gamma)$ is a flattening factor that does not affect the other masses and energy scales reducing only the convexity in the lightest direction and implies a smaller parameter space for models in order to avoid the cosmological moduli problem.
\\
At this point, it is worth saying that the variations that are found are not extremely small and therefore we expect that factors neglected in the analysis, coming from $\ab$ and $3-\gamma$, might alter these results. For example in the actual value for the two bounds: for $\ab<1$ the bounds are underestimated while for $\ab>1$ the bounds are overestimated, compared with numerical results. This, however, at the effective level does not imply a variation of more than a few percents. With a UV approach, even with open string fields in the game, the situation is a bit worst but still, the analysis seems to reproduce quantitatively well the results. In general, numerically we have checked at the effective and UV level, that this overestimates the real shifts for large values of $\gamma$ by almost a factor of two for the extreme case of $\gamma=14/5$ \cite{Gallego2017intrep}.
\\
The flattening effect on the mass is also reflected in the potential barrier height that, naively, it is expected to be of the order of the original VEV for the potential
\be\label{eq:VVEVAdS}
\langle V\rangle_0\sim -\ab \tx^{\ab-1}m_{3/2}^3 M_P\,.
\ee
We now proceed to have a better idea for its value. Using the uplift necessary for a Minkowski vacuum and writing the potential in terms of the minimum critical point,
\be\label{eq:Vcritpoint}
V=\frac{\ab  e^{-\gamma  \varphi } \vp_*^{\ab -1} \left(e^{(\gamma -3) \varphi }-e^{(\gamma -3) \vp_*}\right)}{\gamma -3}+e^{-3 \varphi } \left(\vp_*^\ab-\varphi ^\ab\right)\,
\ee
proceeding like before the idea would be to find a second critical point that is associated to the barrier top. Expanding around $\vp\approx  \vp_*+\epsilon$ we find a solution
\be
\epsilon=\frac{2}{6+\gamma}\,.
\ee
This, however, is not a good estimation for the field excursion to the barrier top is not small enough to neglect higher-order terms in the expansion. Interestingly
\be
h_{bar, rough}\approx \frac12 \partial_{\vp,\vp}^2 V\Big|_{\vp_*}\epsilon^2\approx\frac{2 (3-\gamma )\ab \vp_*^{\ab -1} e^{-3 \vp_*} }{(\gamma +6)^2}\,.
\ee
gives a good value for the barrier height when compared to the numerical results.
\\
Numerical checks show, on the other hand, that the barrier top is, coincidentally, at practically three times the shift just found. Expanding around $\vp_*+6/(6-\gamma)$ we find a much better result for the deviation given by
\be
\Delta \vp=\frac{6}{6+\gamma}+\frac{e^{\frac{18}{\gamma +6}} \gamma  (\gamma +6)+e^{\frac{6 \gamma }{\gamma +6}} (\gamma  (\gamma +18)-90)}{e^{\frac{18}{\gamma +6}} (\gamma +6) \gamma ^2+3 e^{\frac{6 \gamma }{\gamma +6}} (\gamma  (2 \gamma +21)-108)}\approx \frac{30-\gamma }{6 (\gamma +6)}\,
\ee
where in the last equality we approximated the second term to $-1/6$, being a fair value for the range of $1\leq\gamma<3$, where more over $\Delta \vp<0.7$.
\\
This leads to the following expression for the barrier height a leading order in $\vp_*$,\footnote{Another way to get a close estimate for the barrier height is to take the extremal value for the uplift amplitude. There, the potential VEV is approximately twice the value of the barrier height for the uplifting needed to get Minkowski.}
\bea\label{eq:hbarrierTnot}
h_{bar}&\approx& -\frac{\ab  \left(e^{\frac{18}{\gamma +6}} (\gamma +6)+e^{\frac{6 \gamma }{\gamma +6}} (5 \gamma -24)\right) \vp_*^{\ab -1} e^{-\frac{6 (\gamma +3)}{\gamma +6}-3 \vp_*}}{(\gamma -3) (\gamma +6)}\,,
\cr
&\approx & \frac{8 \ab  (3-\gamma ) e^{-3 \vp_*} \vp_*^{\ab -1}}{3 (\gamma +6)^2}\,,
\eea
where again we approximated to an almost identical function in the interval of interest. Plugging back units
\bea
\frac{h_{bar}}{m_{3/2}^3}\sim \frac{8 \ab (3-\gamma )  \vp_*^{\ab -1}}{3 (\gamma +6)^2}M_P\,,
\eea
It is now evident the flattening effect on the barrier height. In particular for $\gamma=14/5$ the factor $(3-\gamma)=0.2$ and together with the other uplift dependent term reaches almost three orders of magnitude lower than the naive expectation from \eqref{eq:VVEVAdS}, a factor that might turn important given the rather low values expected for the volume $\V\sim 10^5$.

\subsection{K\"ahler inflation and Misalignment}\label{sec:misalignmentgen}

Precise statements require a particular inflationary scenario. We choose K\"ahler  inflation \cite{Conlon:2005jm} already studied in \cite{Cicoli:2016olq}. We will review the analysis done there including the effects from the uplift which, given the reduction on the convexity, will lead to a larger misalignment.
\\
In K\"ahler inflation the inflaton role is played by a second blowup modulus that we consider similar in behaviour to the one describe by $x$ in \eqref{eq:LVSpot}. Then, the potential \eqref{eq:LVSpot} is corrected with the following contribution\footnote{An interesting possibility would be to regard the inflaton sector in an ISS setup, like the one considered in \cite{Krippendorf:2009zza}. In this case, the perturbations from the inflaton sector are naturally suppressed, something no always warranted in the standard K\"ahler inflation scenario.} 
\be\label{eq:Vtwoblowups}
V_{inf, UV}=\frac{\lambda_2 e^{-2 y}y^{\beta_2}}{ \V^{2 \eta_2-3}} -\frac{\mu_2 e^{-y} y^{\alpha_2}}{ \V^{\eta_2}}\,,
\ee
with $y$ the inflaton. Both blow up modes are heavier than the volume modulus then it is still possible to integrate them out. Defining 
\be
R=\frac{\lambda  \mu_2^2 (3-\eta_2)^{\ab_2}}{\lambda_2 \mu ^2 (3-\eta )^{\ab }}\,,
\ee
we can simplify again the scalar potential modulo an overall amplitude $\frac{\mu ^2 (3-\eta )^{\ab }}{4 \lambda }$, as
\be\label{eq:inflatondepV}
V=\tF e^{-\gamma  \varphi }+e^{-3 \varphi } \left(\tx^{\ab }-\varphi ^{\ab }-R \varphi ^{\ab_2}\right)
\ee
Here it is possible to notice that the parameter $R$ should be small for the inflaton dynamics do not affect dramatically the stabilization of the other two moduli. In particular, this avoids a large misalignment. Meanwhile, inflation takes place the inflaton dependent exponentials are suppressed, then its contribution to \eqref{eq:inflatondepV} is negligible, in which case we are back to the potential we already studied in previous sections, which is practically flat for the inflaton. Then, we can encode the deformation of the potential due to the inflaton defined in section \ref{sec:cosmohistzeroT}
\be\label{eq:inflndepPot}
V_{inf}= R e^{-3 \varphi } \varphi ^{\ab_2}\,,
\ee
such that the misalignment \eqref{eq:misaligmentgen} takes the form
\be\label{eq:misafull}
\delta \vp\approx\frac{3 R \vp_*^{-\ab +\ab_2+1}}{\ab  (3-\gamma )}\,,
\ee
while the energy associated \eqref{eq:rhomodend} is given by
\be\label{eq:VolumeEnerIni}
\rho_{\vp}^{end}\approx \frac12 m_{\V}^2\delta\vp^2\approx \frac{9 R^2 e^{-3 \vp_*} \vp_*^{-\ab +2 \ab_2+1} }{2 \ab  (3-\gamma )}\,.
\ee
At the this point the factor $3-\gamma$, from the flattening and increasing all these results, should be some how expected.
\\
As pointed out in section \ref{sec:generalmodstab} if the misalignment is too large we could be facing an overshooting. Then, we require
\be
R<\frac{\ab  (30-\gamma) (3-\gamma) \vp_*^{\ab -\ab_2-1}}{18 (\gamma +6)}\,,
\ee
which in general reinforces the constrain $R\ll 1$.

\subsection{Decompactification temperature}\label{sec:thermaleffects}
Before proceeding on including the finite temperature corrections in the analysis let us point out that for these to be relevant for the LVS scenario the temperature must be $T\sim M_P/\V^{3/4}< M_P/\sqrt{\V}\sim M_S$, i.e., it is still lower than the string scale. The same is true for the KK scale, suppressed compared to the previous one, in the geometric regime, by one-fourth of VEV of the blow-up modulus. This is the context we have in mind since we suppose a sector of D-branes, wrapping the blow-up cycle, such that the coupling between the open string modes is controlled by the expectation value of the corresponding modulus. This means that the 4D effective quantum field analysis is consistent and we can proceed by including the corrections \eqref{eq:thermalscalarpotential} with the canonical normalized modulus.
\\
Following the ideas presented in section \ref{sec:modulusthermaleff} we have the shift in the minimum
\be\label{eq:ThermalShift}
\delta_T \phi\approx \frac{\kappa  T^4 e^{3 \vp_*} \vp_*^{-\ab -1}}{\ab  (3-\gamma )}\,,
\ee
to which, from \eqref{eq:rhophireT}, we can associate an energy density given by
\be
\rho^\vp_{re, T}\approx\frac{\kappa ^2 T^8 e^{3 \text{$\varphi $o}} \text{$\varphi $o}^{-\alpha \beta -3}}{2 \alpha \beta  (3-\gamma)}\,,
\ee
and the three different maximal decompactification temperatures depending on the definition: from the barrier top at Minkowski, equation \eqref{eq:Tmaxbargen},
\be\label{eq:TmaxLVSbar}
T_{max,bar}^4\approx \frac{8 \ab  (3-\gamma) \vp_*^{\ab }e^{-3 \vp_*}}{3 (\gamma +6)^2 (\kappa_1 \vp_*-\kappa)}\,; 
\ee
From the saddle point condition, equation \eqref{eq:Tmaxmassgen},
\be\label{eq:TmaxLVSmassT2}
T_{max,mass}^4\approx \frac{\ab  (3-\gamma) e^{-3 \vp_*} \vp_*^{\ab +1}}{(\gamma +6) \kappa }\,;
\ee
And from the shift one, equation \eqref{eq:Tmaxshiftgen},
\be\label{eq:TmaxshiftLVS}
T_{max,shift}^4\approx \frac{6 \ab  (3-\gamma) e^{-3 \vp_*} \vp_*^{\ab +1}}{(\gamma +6) \kappa }\,.
\ee  
As advertised, among the results \eqref{eq:TmaxLVSmassT2}, \eqref{eq:TmaxshiftLVS} and \eqref{eq:TmaxLVSbar} the first one is the most restrictive, i.e., the barrier top will be at negative values before getting a saddle point or that the shift is comparable to the distance to the barrier top, something somehow expected. Choosing this as our decompactification limit and restoring units to compare to other scales in the game we have
\be
T_{max}\approx \left(\frac{8 \ab  (3-\gamma) \vp_*^{\ab }}{3 (\gamma +6)^2 (\kappa_1 \vp_*-\kappa)}\right)^{1/4}\left(\frac{m_{3/2}}{M_P}\right)^{3/4}M_P\,.
\ee

\subsection{LVS, K\"ahler inflation and modular cosmology}

We now study the particular situation of LVS with K\"ahler inflation reviewing results in \cite{Cicoli:2016olq} following the general discussion presented in section \ref{sec:cosmohistzeroT}.\footnote{Although a misalignment for the $x$ mode is also possible we will simply neglect this for two reasons: the first one is the evident decoupling between the inflaton and the $x$ modulus; the second one is the fact that the Hubble scale during inflation and the mass of the inflaton is suppressed compared to the one of the $x$ mode by a factor of $R$ making that any deviation from the minimum settles down quickly before the decay of the inflaton.} We will make the approximation of an instantaneous reheating, i.e., $N_{re}=0$ and $\rho_{re}=3M_P^2 \Gamma_\tau^2$, motivated from the results in \cite{Barnaby:2009wr} predicting a violent preheating phase.

\subsubsection{Inflaton domination}
The initial inflaton domination comes from a violent and efficient production of inflaton quanta lasting until its decay. The general expression for the number of e-folds is given in \eqref{eq:taudomEfold} and for our case this turns to be
\be\label{eq:NtauLVS}
N_\tau\approx\frac13\ln\left(\frac{100}{3}\frac{ e^{ \varphi _*}}{R^2 \varphi_* ^{2\ab_2+1/2}} \right)\sim \frac{\vp_*}{3}+1
\ee
where we used \eqref{eq:inflndepPot} and $\Gamma_\tau\approx 0.1 m_\tau^2 \V/ M_P^2$ \cite{Cicoli:2010ha} such that
\be
\Gamma_\tau \simeq \frac{4 e^{-2 \vp_*} }{15 \sqrt{3}}\left(\frac{a_2^{3/2} \sqrt{3-\eta_2} \mu ^2 R (3-\eta )^{\ab } \vp_*^{\ab_2+\frac{1}{2}}}{\alpha _2 \lambda  \lambda_2}\right)^{3/2} \sim 0.1 e^{-2 \vp_*} R^{3/2} \vp_*^{\frac34(2\ab_2+1)}\,,
\ee
where we omitted numerical factors like the overall $\frac{\mu ^2 (3-\eta )^{\ab }}{4 \lambda }$ in the mass also neglected in the scalar potential. 

\subsubsection{Radiation dominated}

Given the definition \eqref{def:epsilon} from \eqref{eq:inflndepPot} and \eqref{eq:VolumeEnerIni}
\be
\epsilon^2=\frac{9 R \vp_*^{-\ab +\ab_2+1}}{2 \ab  (3-\gamma)}\,,
\ee
the general result \eqref{eq:Nrad2} leads to 
\be
N_{rad}=-\ln\left(\frac{9 R \vp_*^{-\ab +\ab_2+1}}{2 \ab  (3-\gamma)}\right)\,.
\ee
This as far no finite temperature effects are taken into account. A proper consideration of such effects leads to the general result \eqref{eq:NradT} with the definition \eqref{def:theta}. In our particular case with \eqref{eq:misafull} and  \eqref{eq:ThermalShift}, while $e^{-N_\tau}\approx3 M_P^2\Gamma_\tau^2/V_{inf}$ 
\be\label{eq:thermalvsnonthenergies}
\theta^2\approx \frac{ R^4  \vp_*^{4\ab_2+11/2}}{ \kappa ^2 \tT^8 }e^{-\vp_*}\,.
\ee
using\footnote{A precise expression for the reheating temperature, from $\rho=\frac{\pi}{30}g_*T^4_{rh}=3 M_P^2\Gamma_\tau^2$ leads to a $\tilde T\sim e^{\vp_*/4}$ which seems to invalidate the following assumption. However, for well-motivated values of the volume from K\"ahler inflation, i.e., $e^{\vp_*}\sim 10^5-10^7$, the difference to the assumed value can be comparable to the numerical factors anyway disregarded. For simplicity, we thus take $\tilde T$ as something not scaling with the volume since the conclusions do not change dramatically.}
\be
\tilde T= e^{3\vp_*/4}T\simeq{\cal O}(1)\,,
\ee 
such that
\be
N_{rad,T}\approx\frac13 \ln\left( 0.1\frac{\ab R^3 (3-\gamma)  \vp_*^{3\ab_2+\ab +9/2}}{\kappa ^2 \tT^8}e^{-\vp_*}\right)\,.
\ee
Here we see explicitly how thermal corrections reduce the number of e-folds during the first radiation dominated epoch by inducing a larger modulus energy density. In fact in the present case seems even that at our level of approximation this epoch is absent as naively leads to $N_{rad,T}\approx -\vp_*/3<0$. From the general discussion in section \ref{sec:modulusthermaleff} we infer that most likely we must consider details of the numerical factors to get a proper conclusion.

\subsubsection{Volume modulus domination}

From the usual decay rate for fields interacting with gravitational strength, $\Gamma\approx\frac{1}{16\pi}m^3/M_P^2$ and given the canonical normalized volume modulus $m_\vp^2\sim e^{-3\vp_*/2} M_P^2/\vp_*$ we are dealing with a very long-lived scalar field which, however, can be heavy enough to avoid the moduli problem. In particular, for well-motivated values of the volume in a K\"ahler inflation scenario, this modulus is not lighter than $10^9GeV$, far above the constraints from BBN. Still, its late decay implies a second modulus domination, besides the one from the inflaton.
\\
The general expressions for the number of e-folds, \eqref{eq:NeModdomGen} and \eqref{eq:NmodT},  lead to  
\be\label{eq:NphiLVSnoT}
N_\vp\approx\frac53\vp_*+\frac23 \ln\left(\frac{200 R^{7/2} \vp_*^{-\frac{1}{4}(8\ab -14\ab_2-17)}}{ (3-\gamma)} \right)\,,
\ee
if thermal effects are neglected or
\be\label{eq:NphiLVSfiniteT}
 N_{T,\vp}\approx 2\vp_* +\frac13 \ln\left(\frac16\kappa^2 \tilde T^8\right)\,,
\ee
in these are considered.

\subsubsection{Final reheating temperature}

From the general results \eqref{eq:FinalTnThermal} and  \eqref{eq:FinalTwThermal} we get, with $g_*\sim 100$,
\begin{equation}
T_{rh}\approx \frac{1}{4 \pi } \tilde\alpha ^{3/4} (3-\gamma )^{3/4} \vp_*^{3 (\tilde\alpha -1)/4} e^{-9 \vp_*/4}\,,
\end{equation}
and
\begin{equation}
T_{rh,T}\approx \frac{\kappa ^2}{4 \pi R^4} \tilde\alpha ^{3/4} (3-\gamma )^{3/4} \vp_*^{ (3\ab-16\ab_2 -25)/4}  \tT^8  e^{-\vp_*/4} \,.
\end{equation}
Here it is completely evident how thermal corrections could help in increasing the final reheating temperature and therefore in alleviating the tension from the moduli problem.

\subsubsection{Preferred window for inflation e-folds}

We now implement our general result \eqref{eq:numEfolds} (see also \cite{Dutta:2014tya,Das:2015uwa}) to the case of LVS and K\"ahler inflation, also worked out for the zero temperature case in \cite{Cicoli:2016olq}. K\"ahler inflation, being a kind of Starobinski scenario, leads to a very small tensor to scalar ratio, $r=16\epsilon_{sl}\sim 10^{-10}-10^{-11}$ with $\epsilon_{sl}$ the slow-roll parameter, for well motivated values of the volume $\V\sim10^5-10^6$ \cite{Cicoli:2016olq}. Then, with a negligible contribution from the ratio $\rho_k/\rho_{end}$, in this case we have
\be
N_k\approx50-\frac14\left(N_\vp+N_\tau\right)\,.
\ee
Using \eqref{eq:NtauLVS} and \eqref{eq:NphiLVSnoT}, neglecting other numerical factors,
\be
N_k\approx 50-\frac12\vp_*\,,
\ee
while from the one rising from thermal corrections \eqref{eq:NphiLVSfiniteT}
\be
N_k\approx 50-\frac{7}{12}\vp_*-\frac{1}{12}\ln\left(\frac16\kappa^2 \tilde T^8\right)\,.
\ee
Now, with well motivated values for the K\"ahler inflation scenario $\vp_*\sim 10$, we have $N_k\approx 45$ without thermal corrections and $N_k\approx 44$ in presence of this finite temperature effects.
\\
As was pointed out in \cite{Cicoli:2016olq} this deviation from the standardly accepted value of $N_k\approx 50-60$ might lead to differences in the observables, for example, the scalar tilt in K\"ahler inflation takes the simple form \cite{Conlon:2005jm}
\be
n_s\simeq 1-\frac{2}{N_k}\,,
\ee
that in near future could be pinpointed experimentally. We stress again that, not only the presence of finite temperature effects might induce larger corrections to the ones expected from the initial misalignment, but also that these can be present even in the case the misalignment turns to be null for some reason.

\section{Instabilities from oscillon production}\label{sec:Floquetan}

In this report we have explored, first with some generality and then with the explicit example of LVS, how finite temperature effects, through corrections to the scalar potential might induce a different cosmological history in presence of light moduli.
\\
We close our discussion rising a further possibility, namely the
production of oscillons: localised, long-lived, non-linear excitations of the scalar fields; this strongly depends on the shape of the scalar potential, and their presence might affect the cosmological evolution as they can dominate the energy density and delay the final reheating \cite{Kane:2015jia,Copeland:1995fq}. For these, we present explicit partial numerical examples for LVS showing that thermal corrections potentially enhance the possibility of oscillons in this kind of vacua.
\\
The conditions for oscillon production are well-known \cite{Amin:2010dc,Amin:2013ika} with main features resumed as: first, perturbations around the oscillatory homogeneous solution grow and this grow is strong enough for the non-linear interactions to become important; and second, around the minimum, the potential is shallower than $\phi^2$, at least in the region where the dynamics are important. From the second condition, we see how finite temperature corrections might in general play a role favouring oscillon production since these decrease the convexity of the potential. Still, definite conclusions need a numerical study in a case by case basis. In \cite{Antusch:2017flz} the generation of oscillons in string compactification was studied performing the full numerical analysis. Here we restrict ourselves to the Floquet analysis for the evolution of the scalar fluctuation, indicating the possibility of a phase of rapid growth. This is expressed in an instability diagram where the so-called Floquet exponents are plotted, and whose size express the possibility of strong dynamics. 

\subsection{Floquet analysis and exponents}

Oscillons can be formed if the quantum fluctuations $\delta \phi(t,\vec x)$ around the homogeneous background $\phi(t)$,
\be
\phi(t,\vec x)=\phi(t)+\delta \phi(t,\vec x)\,,
\ee
grow exponentially due to some resonance effect. For the homogeneous part, we take a zero-order approximation by neglecting the universe expansion. This allows two things in our analysis: first, the oscillation for the volume modulus is completely periodic\footnote{We regard also time scale shorter than the meantime life so that we can neglect the decay rate as well.} and second, the temperature will be constant in our calculations. This is, of course, a crucial detail that however, to our knowledge, would be anyway hard to implement in a full-fledged numerical analysis. Having a periodic behaviour we can tackle the problem using a Floquet analysis, whose details are given for completeness in appendix \ref{app:floquet} but for further discussion, we refer to \cite{Amin:2011hu}.
\\
Working in momentum space the e.o.m. for the fluctuation takes the form
\be
\frac{\partial x(t)}{\partial t}= E(t)x(t)\,,
\ee	
where
\be
x=
\left(
\begin{array}{c}
	\delta \phi _k \\
	\partial_t \delta \phi _k \\
\end{array}
\right)\,,~~\text{and}~~
E(t)=\left(
\begin{array}{cc}
	0 & 1 \\
	-k^2-V''( \phi)  & 0 \\
\end{array}
\right)\,.
\ee
Then, via the Floquet theorem, the solution can be stated as
\be
x(t) \propto  e^{\pm \mu (t-t_0)}\,,
\ee
with $\mu$ the so-called Floquet exponent, which more commonly are evaluated through the Floquet multipliers $\pi_\pm =e^{\pm T\mu}$ with $T$ the period of oscillation, such that
\be
\mu =\frac{1}{T}\left(\ln\left(\left|\pi _+\right|\right)+\rmi  \text{Arg}(\pi _+)\right)\,.
\ee
Therefore, an exponential grow appears whenever $\Re(\mu)\neq0$ or $\left|\pi _\pm\right|\neq0$. Floquet multipliers can be evaluated from the linear independent solutions to the e.o.m. $x_1(t_0 )^T=\left(1,0\right)$ and $x_2(t_0 )^T=\left(0,1\right)$, such that
	\be
\pi_{\pm}^k=\frac{1}{2}\left.\left(x_1^1+x_2^2\pm\sqrt {\left[x_1^1-x_2^2\right]^2+4 x_1^2x_2^1}\right)\right|_{t=t_0+T}\,.
\ee

\subsection{LVS finite temperature}
\begin{figure}[htb]
	\begin{center}
		\includegraphics[width=0.7\textwidth
		]{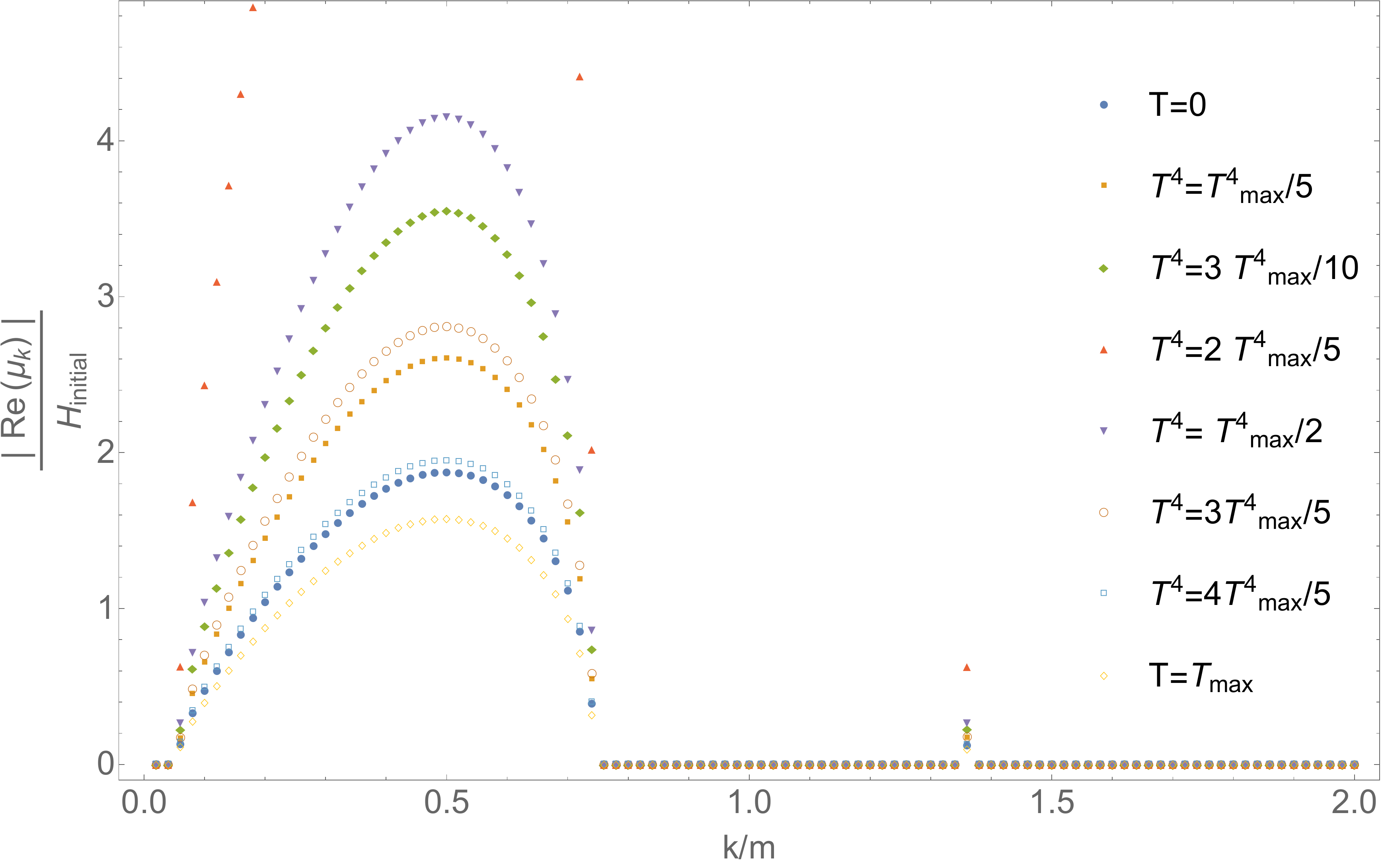}
	\end{center}
	\caption{Normalized real part of the Floquet exponents for the LVS scalar potential at finite temperature for $\gamma=14/5$. The initial value of the amplitude is taken to be $\phi_{in}=\phi_*+(\phi_{top}-\phi_{inf})/2$, the subscripts denoting the barrier top position and the inflection point. Very large values, up to order $10$, are obtained for values near $T^4\approx \frac25 T_{max}^4$, with a maximum value of $|Re(\mu_k)|/H_{in}\approx 9.7$, after which the values are roughly of the same size to the ones for zero temperature.}\label{fig:floquetLVST} 
\end{figure}

\begin{figure}[htb]
	\begin{center}
		\includegraphics[width=0.7\textwidth
		]{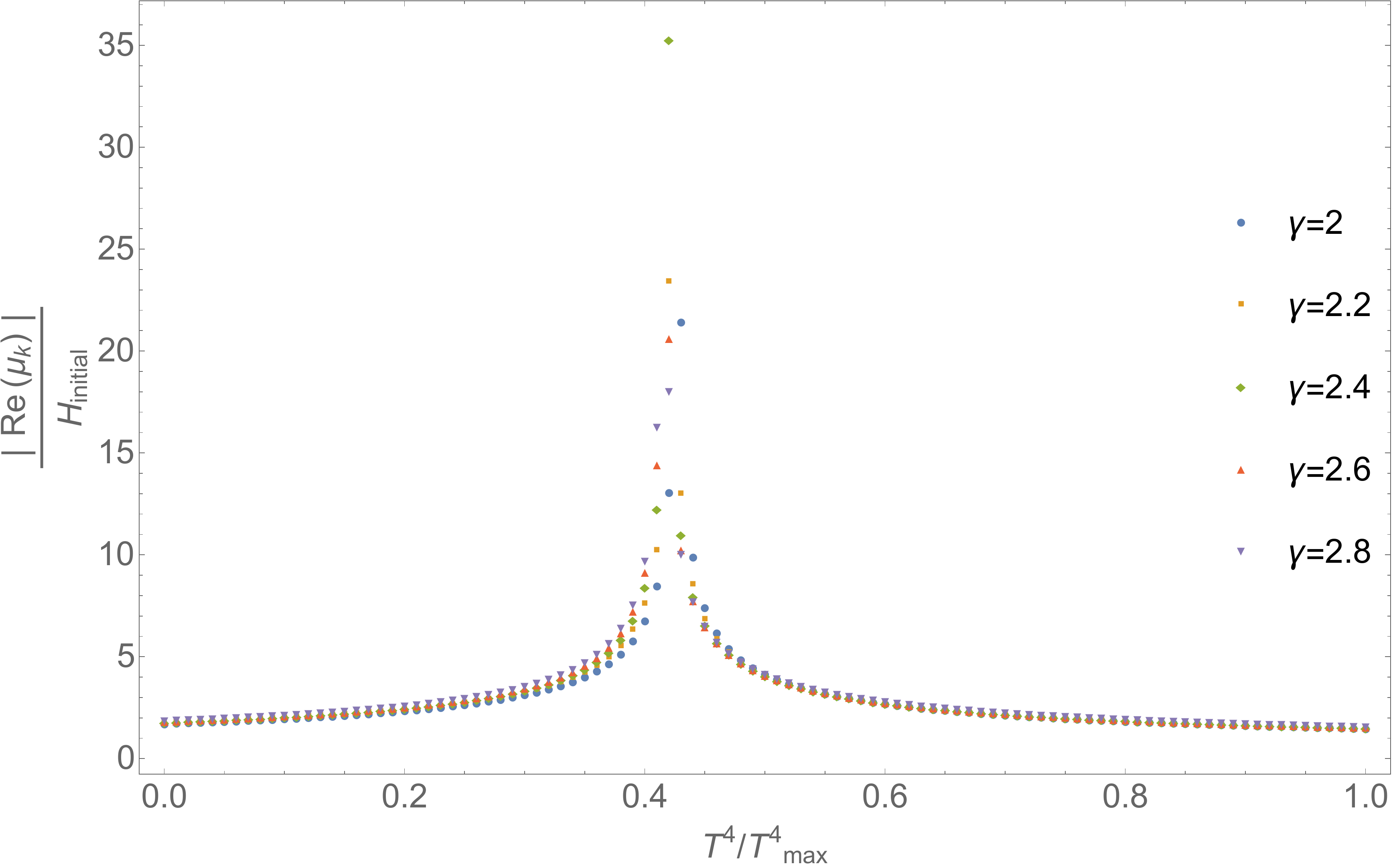}
	\end{center}
	\caption{Maximum value for the Floquet exponents as a function of the temperature. We notice a divergent behaviour around $T^4/T_{max}^4\approx 2/5$ for all values of uplift parameter $\gamma$.}\label{fig:floquetmax}
\end{figure}

\begin{figure}[htb]
	\begin{center}
		\includegraphics[width=0.7\textwidth
		]{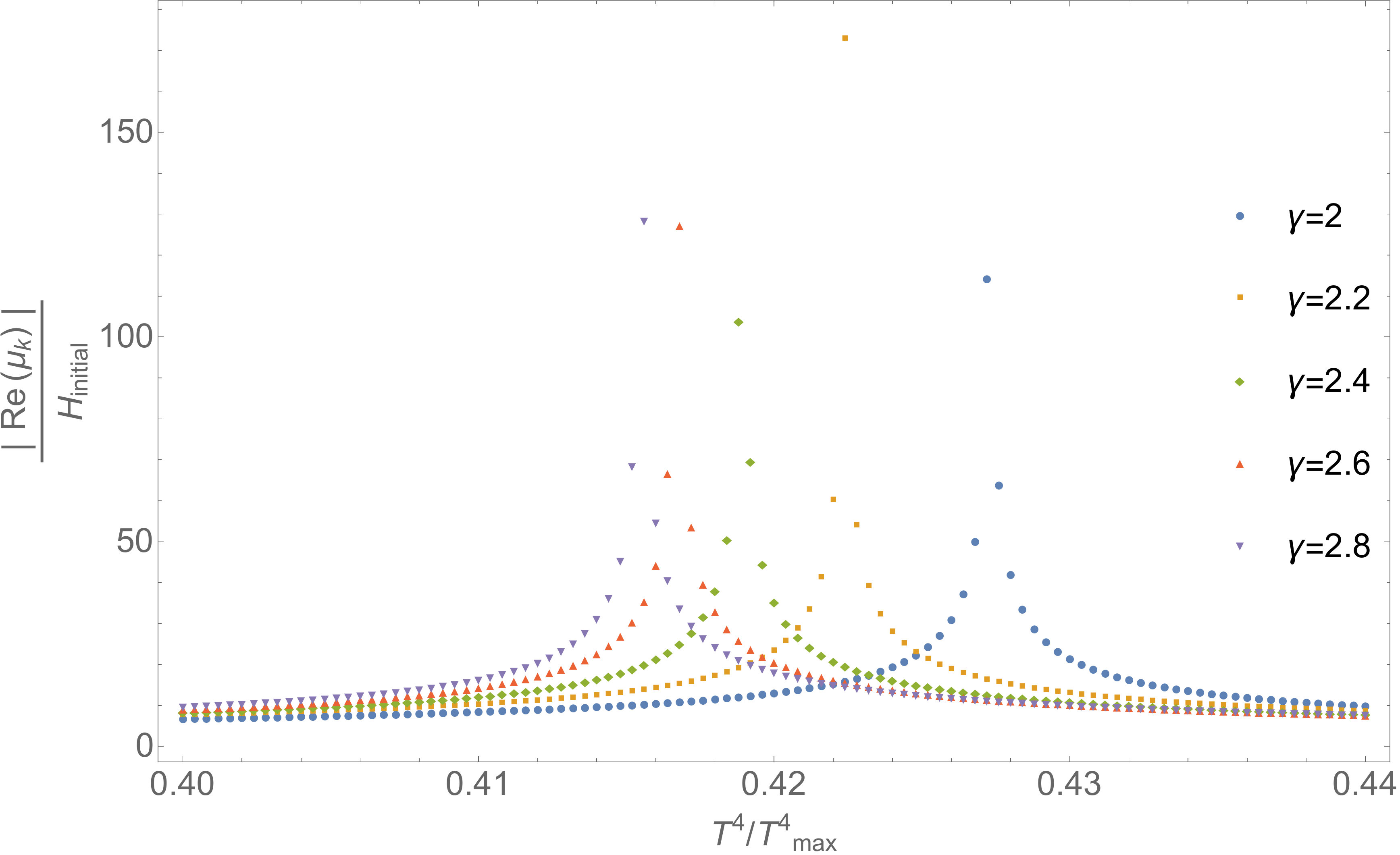}
	\end{center}
	\caption{A detail of the previous plot for maximum value for the Floquet exponents as a function of the temperature. The divergent points do not coincide  for all $\gamma$'s. Here we confirm a divergent behaviour being quite strong the increase making imposible, with our resolution, to spot if there is a finite maximal value.}\label{fig:floquetmaxdetail}
\end{figure}

For this analysis we use the potential as appears in \eqref{eq:Vcritpoint} with the canonical normalized field $\phi=\sqrt{\frac23}\vp$. We change the temperature for different values of the uplift exponent (the zero temperature study varying the uplift is shown in appendix \ref{app:LVSOsciup}). 
\\
We chose $\tilde \alpha=1$ and the value of the position of the  potential minimum, though irrelevant for the analysis, is taken to be $\phi_*=5$. For the finite temperature correction we take $\kappa=1$ and $\kappa_1=5$, and fixing the uplift parameter to the extreme case $\gamma=14/5$, we vary the temperature in factors of the maximal temperature \eqref{eq:TmaxLVSbar}. In all cases, we take for the initial amplitude the midpoint between the barrier top and the inflection point. The results are shown in figure \ref{fig:floquetLVST}.
Naively it is expected that the increase of the temperature, and subsequent flattening of the potential, would induce further instabilities in the system, revealed in an increase in the Floquet exponents.  This is precisely what is seen in the plot but until a critical temperature is reached. Afterwards, the exponents start to decrease with an increasing temperature, reaching similar values to the ones of the zero temperature case. We have checked that the situation is not changed for different values of $\gamma$ or minimum point $\phi_*$. We also checked this for different values of the uplift exponent with, interestingly, a very similar critical temperature.
In order to explore better this behaviour we concentrate on the maximal value, located at $k/m\approx1/2$ and plot them as a function temperature. This is shown in figure \ref{fig:floquetmax} with a detailed region in figure \ref{fig:floquetmaxdetail}.

The situation clearly reveals no only an unexpected behaviour after an almost universal critical temperature. At this critical temperature, moreover, the exponents seems to explode. Unfortunately we do not have ideas on how to explain this behaviour nor the numerical tools to refine our analysis for a complete numerical study which will definitely reveal possible oscillon generation.

\section{Conclusions}

We explored how finite temperature corrections can affect in general not only the stabilization of light moduli but also the cosmological evolution of the universe. In particular, we noticed that thermal effects in the scalar potential might, in general, induce a sort of a kick in the modulus oscillation from the shift on its minimum just after a first reheating from the inflaton decay. This energy imprint on the modulus changes the subsequent evolution and increases the final reheating temperature with possible important cosmological implications for model construction. This is also reflected in observables like the scalar tilt, $n_s$, through the preferred number of e-folds for inflation. An important observation is the fact that these effects are likely to be more important to the ones coming from the initial misalignment since the thermal kick happens at the end of the reheating epoch, a time at which the initial energy in the modulus has been already redshifted. These ideas are shown in general in section \ref{sec:ThermalSec} after reviewing, in section \ref{sec:generalmodstab}, the well-known case at zero temperature. These general results are then explicitly illustrated for Large Volume Compactifications in a K\"ahler inflation scenario confirming previous expectations but also revealing the need for a more detailed numerical analysis in order to extract precise conclusions.
\\
A central point in our work is the study of possible consequences due to the finite temperature deformations of the scalar potential. For this, we include in the scalar potential the deformation given in \eqref{eq:thermalscalarpotential} regarding that the modulus under study plays the role of the coupling entering in the loop computations. Physically this implies a coupling between the thermal bath and the modulus, something that is in general expected for closed string moduli from superstring compactifications, and even for open string ones is not difficult to imagine such. These corrections thus are generally present and precise statements regarding, for example, modular cosmology should keep these into account.
\\
An initial worry in this kind of studies is the possibility of a decompaction due to the thermal correction and several works studied such a possibility in several scenarios \cite{Buchmuller:2004xr,Buchmuller:2004tz,Anguelova:2007ex,Anguelova:2009ht} although numerical analyses seem to show that these are probably too much conservative \cite{deCarlos:1993wie}. We revisit some of the arguments behind a decompactification temperature pointing out that in general there might be at least three ways of defining such but all leading to results of the same order of magnitude for generic scalar potentials. Still, the one usually used, simply regarding that, the barrier height is equal to the absolute value of the thermal scalar potential, turns to be also the most conservative estimation, something that might explain the negative results from the numerical studies.
\\
The main point in the whole report is to notice that one of the effects of thermal corrections is a shift of the minimum compared to the zero temperature one (see also \cite{Nakayama:2008ks}). Then, in a similar way to the misalignment present due to the coupling between the moduli and the inflaton, the moduli oscillation is affected and some of the energy is injected into the modulus. However, since finite temperature effects appear just after a thermal bath has been formed and this happens only after reheating, the energy from the misalignment has been already redshifted and it is most probably subleading. More precisely the energy stored in the modulus after reheating is larger by a factor $\theta^{-2}=3 M_P^2 \Gamma_\tau e^{3N_{rad}}/V_{inf}>1$ compared to the energy at the same time if thermal effects are neglected.
\\
This last observation implies that the moment of radiation/matter equality comes earlier and the late modulus dominated epoch is longer. Two automatic consequences are that the final reheating temperature is higher by a factor $\theta^{3/4}$, possibly alleviating potential tensions with BBN; secondly, in order to ensure a good timing for horizon re-entry the number of e-folds from horizon exit and the end of inflation should be smaller than usually though. A shorter period of inflation is a well known fact of modular cosmology and in general is shortened by a term $\frac14 N_{mod}$ \cite{Dutta:2014tya}. Then, by including thermal corrections the number of preferred e-folds is shortened by a factor of $\frac{1}{12}\ln\left(\theta^2\right)$.
\\
A precise estimation of the preferred window of inflationary e-folds was done for the particular case of LVS in a K\"ahler inflation scenario \cite{Cicoli:2016olq}. They find in particular that the value is shortened by a term $-\frac12 \ln(\V)$ while introducing thermal corrections, the factor $\theta^2\sim \V^{-1}$, and the term shortening this value is given by $-\frac{7}{12}\ln(\V)$. Indeed it is just a small difference but the important fact is that this last result is larger and present even in the case that the misalignment is absent, but of course, if thermal corrections to the scalar potential take place.
\\
Overall our results might imply modification in studies like the ones done in \cite{Bhattacharya:2017ysa} for the reheating equation of state parameter or in  \cite{Bhattacharya:2017pws} to constrain the K\"ahler inflation model. This is also the case for detailed analyses on the present matter abundance like the one done in \cite{Acharya:2019pas,Allahverdi:2019jsc} where the energy stored in the moduli is regarded, as usual, as coming exclusively from the misalignment. In particular, the thermally induced energy injection into light moduli components makes the contribution from heavier ones further irrelevant, and dark matter overproduction is possibly controlled without the need of a dark radiation portal.
\\
On the other hand scenarios with scalar fields, like the ones considered in \cite{Maharana:2017fui,DiMarco:2018bnw} would in principle suffer from these modifications but in these cases justifying a coupling with the thermal bath would not be an easy task for these scalars are supposed to be decoupled, or weakly coupled, to the observable sector and dark radiation constraints would probably forbid a situation like the one discussed here. However, the situation can be generically extended to any analysis concerning the cosmological evolution, like the one done in \cite{Amin:2019qrx} and it would be interesting to consider more exotic scenarios like the ones handling with hidden thermal baths \cite{Hardy:2019apu}.
\\
One natural complaint in our approach is that in general, we have taken for granted that the thermal kick in the moduli oscillation really takes place and the shift in the minimum does not occur in an adiabatic way. We expect to proceed in the future to better establish the condition under which this kick is really present and the effective temperature entering on it \cite{WithFrancescoandAnshuman}. This would probably imply numerical studies similar to the ones done in \cite{Barreiro:2007hb}.
\\
A further possible consequence from the scalar potential deformation and more precise the decrease in its convexity due to the thermal corrections is a rise in the oscillon production and therefore changes in the cosmological evolution. In section \ref{sec:Floquetan} we study this possibility for LVS compactifications already studied for the zero temperature case in \cite{Antusch:2017flz}. We do this
through a Floquet analysis varying the temperature and evaluating the Floquet exponents to spot possible instabilities. We find that for low temperatures an increase on it is reflected in larger exponents, something expected due to the just mentioned decrease in the convexity, but interestingly enough after some critical temperature, the exponents start to decrease until their value is close to the zero temperature case. To this, for us estrange, behaviour it is summed the fact that the critical temperature is rather universal and close to $T^4_{crit}\approx\frac25 T^4_{max}$ and that at this point the Floquet exponents seem to present a divergent behaviour, in a kind of a lambda phase transition diagram. Despite this last observation might be a clear indication of instabilities, a proper oscillon production analysis requires a complete numerical study, and lattice simulations. Unfortunately, we do not have the technical capabilities for doing such but we hope this serves as a motivation to pursue this kind of studies, including thermal corrections either in the instability analysis or in more general contexts like the study of compact objects from scalar theories \cite{Muia:2019coe}, with interesting cosmological consequences like primordial black holes \cite{Martin:2019nuw}, primordial gravitational waves \cite{Dufaux:2007pt,Assadullahi:2009nf,Antusch:2017vga,Kohri:2018awv,Marsh:2015wka} , and even a reheating scenario through Hawking radiation \cite{Lennon:2017tqq}.\footnote{I thank Francesco Muia for drawing my attention to these references and the possible consequences of my studies.}
\\
All these ideas come as small contribution to the  join efforts to put forward phenomenological consequences that could be experimentally tested in near future \cite{Abazajian:2016yjj,Finelli:2016cyd} and hopefully set ground for a renewed interplay between theory and experiment.

\acknowledgments
I would like to thank Anshuman Mahanara, Francesco Muia, Fernando Quevedo, Yeinzon Rodríguez and Timm Wrasse for fruitful discussions, and Mustafa Amin for email exchange and guidance with the code for the Floquet analysis. Special thanks to Anshuman Mahanara and Francesco Muia for comments on a preliminary version of the manuscript. I also acknowledge the hospitality at the HECAP section of ICTP Trieste, the Harish-Chandra Research Institute, the Grupo de Investigación en Relatividad y Gravitación from Universidad Industrial de Santander and the Institute for Theoretical Physics, TU Wien, where the ideas in this work were born or partially done.
\appendix
\section{Explicit LVS models and effective parameters}\label{app:explicitUVLVS}
Here we briefly present the standard LVS model with only moduli fields and one with open string to get explicit values for the parameter appearing in \eqref{eq:LVSpot}.
\subsubsection{No open strings}
In this case besides the K\"ahler potential in \eqref{eq:Kalerpotgen} the non-perturbative component for the superpotential  in \eqref{eq:Wgen} is given by
\be
W_{np}= \sum_i A_i e^{-a_i t_i}\,,
\ee
with the factors $a_i$ is $2\pi$ for Euclidean D3-branes and $2\pi/N_{\rm D7}$ for a stack of $N_{\rm D7}$ branes. The amplitudes $A_i$ carry possible axio/dilaton and complex structure moduli dependencies, which however are not relevant for us here for we consider them as fixed.
\\
With these consideration in the SUGRA scalar potential \eqref{eq:FtermPot} regarding only leading terms in powers of the volume and the exponentials with $a_i t_i>1$,
\bea\label{eq:LVSpotential}
V_{\rm LVS} &=& e^{\cal K} \left[
\frac{3 \tilde\xi }{2\V^3 } |W_{flux}|^2 
+\sum_i \left(\frac{8\sqrt{t_i}}{3\tilde\kappa^{3/2}_i\V} a_i^2 |A_i|^2 e^{-2a_i t_i} - 
	\frac{4  a_i t_i|A_i W_{flux}|}{\V^2}   e^{-a_s t_i} 
\right)
\right] \,,
\, ,
\eea
with $\tilde \kappa_i$ the final proportional factor in $\V=\kappa_T T^{3/2}-\sum_i \tilde \kappa_i t^{3/2}_i$. We have also chosen the axionic components of the moduli to be such that the relative signs turn in this way warranting a minimum in these directions at vanishing VEV. Comparing with the expression \eqref{eq:LVSpot} we have
\bea
\eta&=&2\,,~~~\zeta= \frac32 e^{\cal K}\tilde\xi  |W_{flux}|^2\,,\\
\alpha&=&1\,,~~~ \lambda_i=\frac{8a_i^{3/2}}{3\tilde\kappa^{3/2}_i} e^{\cal K}|A_i|^2\,,
\\
\beta&=&\frac12\,,~~~ \mu_i=4 e^{\cal K} |A_i W_{flux}|\,.
\eea
\subsection{Open strings}
The presence of open strings, in general, introduces chiral moduli charged under a $U(1)$ local symmetry which in turn is pseudo anomalous, with the anomaly cancelled via a Green-Schwarz mechanism that induces the non-linear charge in the moduli. We consider the case where is the small modulus the charged one. 
\\
More properly the situation is a non-abelian group $SU(N_c)$ with $N_f$ flavours, and we stick to the case $N_f<N_c$ as proposed in \cite{Cremades:2007ig} and studied in \cite{Gallego:2011jm}. The chiral fields $Q$  ($\tilde Q$) transform in the fundamental (antifundamental). However, it is possible to work with mesonic degrees of freedom $\Phi=\sqrt{2 Q\bar Q}$ characterising D-flat directions. 
\\
Within our working window an $ADS$ nonperturbative superpotential is
generated \cite{Affleck:1983mk,Affleck:1984xz}, depending on the non-perturbative scale $\Lambda$ that the depends on the closed moduli, controlling the couplings, and the mesonic field:
\begin{equation}
W_{np}=(N_c-N_f)\left(\frac{2\Lambda^{3N_c-N_f}}{\Phi^{2N_f}}\right)^{\frac{1}{N_c-N_f}}\,.
\end{equation}
In order to get simpler expressions we take the explicit case with $N_c=2$ and
$N_f=1$, then the superpotential takes the general form
\begin{equation}
W_{mod}=-A \frac{e^{-a\, t}}{\Phi^2}-\frac12  m \rho \Phi^2\,,
\end{equation}
where we introduced a mass term for the mesonic field, which requires a further charged field, $\rho$, singlet under the $SU(2)$. The phases are given by the stabilization of the axionic components in minima of the potential with zero VEV.
\\
Without loss of generality we normalized the charges such that the one of the modulus is such that $\delta^t=2/a$, so the holomorphic Killing vectors associated to all four fields are given by $X_X=i(0,1/a,-\phi/2,\rho)$. The K\"ahler potential for the matter
fields is taken  to be $K\supset \frac{Z(t)}{T^n}|Q|^2$, with some modular weight $n$, so
the leading expression for the $D$-term
\begin{equation}
D_X=-iX^iK_i\approx\frac12\left(\frac{1}{a}\frac{\tilde \kappa_t}{\kappa_T}\frac{3\sqrt{t}}{T^{3/2}}+\frac{Z(t)}{T^n}(2\rho\bar\rho-\phi
\bar \phi)\right)\,,
\end{equation}
where we have discarded subleading terms in $1/at$.
\\
The dynamics are dominated by the D-term potential then minimization leads to a
nearly cancellation of the $D$-term via a non vanishing VEV for
$\phi$, $\langle \phi^2\rangle\approx
\frac{3\tilde \kappa_t\sqrt{t_r}}{a\kappa_TZ(t_r)T_r^{3/2-n}}\equiv\frac{\phi_o(t_r)^2}{T^{3/2-n}}$, taking $\langle \phi \rangle\gg \langle \rho\rangle$. The uplifting term in this scenario might come from F-term dynamics of the $\rho$ fields and require $m\sim T^{\frac32-\frac94n}$. This induces a size for $\rho$ at the minimum at most of order $\rho \sim T^{\frac34n-\frac52}$. All these, with natural values for $n$ to avoid Planckian order dynamics, imply that we can at first order take $\rho=0$ (for more details see \cite{Gallego:2011jm}). The moduli scalar potential then read at leading order
\bea\label{eq:LVSpotentialopenstrings}
V_{\rm LVS} &=& e^{\cal K} \left[
\frac{3 \tilde\xi }{2\V^3 } |W_{flux}|^2 
+\frac{8 a^4 Z(t)^2 \kappa_T^{4n/3} |A|^2}{27 \kappa _t^3 \sqrt{t} \V^{\frac{4}{3} n-1}}    
e^{-2a t} -\frac{4 a^2 \sqrt{t} |A W_{flux}| Z(t)\kappa_T^{2n/3}   }{3 \kappa_t  \V^{\frac{2}{3} n+1}} e^{-a t} 
\right] \,.
\eea
Then taking $Z(t)= c t^\ell$ we identify,
\bea
\eta&=&\frac23 n+1\,,~~~\zeta= \frac32 e^{\cal K}\tilde\xi  |W_{flux}|^2\,,\\
\alpha&=&\ell+\frac12\,,~~~~~\lambda=\frac{8 c^2 a^{\frac{9}{2}-2 \ell}\kappa_T^{\frac{4}{3}  n} |A|^2e^{\cal K}  }{27 \kappa_t^3} \,,\\\cr
\beta&=&2\ell-\frac12 \,,~~~~\mu=\frac{4 c  a^{\frac{3}{2}-\ell} \kappa_T^{\frac{2}{3}  n} |A W_{flux}| e^{\cal K} }{3 \kappa_t}\,.
\eea

\section{Floquet analysis} \label{app:floquet}
\subsection{Floquet analysis and exponents}
Here we resume a procedure whose details can be found in \cite{Amin:2011hu}. Oscillons might appear after an exponential growth of the fluctuation around the homogeneous solution
\be
\phi(t,\vec x)=\phi(t)+\delta \phi(t,\vec x)\,.
\ee
For the homogeneous part we disregard the universe expansion, solving
\be
\ddot{\phi}+V'(\phi)=0\,,
\ee
while for the fluctuation we work in the momentum space
\be
\delta \phi=\int \frac{d^3k}{(2\pi)^3}\delta \phi_ke^{\rmi \vec k\cdot \vec x}\,,
\ee
satisfying
\be\label{eq:dletaphieom}
\partial _t^2\delta\phi _k+\left(k^2+V''(\phi)\right)\delta \phi _k=0\,.
\ee
The important thing to notice is that, given that we neglect the universe expansion, the coefficients are periodic in time and Floquet theory can be applied. This equation is more elegantly written as
\be
\frac{\partial x(t)}{\partial t}= E(t)x(t)\,,
\ee	
where
\be
x=
\left(
\begin{array}{c}
	\delta \phi _k \\
	\partial_t \delta \phi _k \\
\end{array}
\right)\,,~~\text{and}~~
E(t)=\left(
\begin{array}{cc}
	0 & 1 \\
	-k^2-V''( \phi)  & 0 \\
\end{array}
\right)\,.
\ee
Defining the fundamental matrix $O$ by
\be
\frac{\partial O\left(t-t_0\right)}{\partial t}=E(t)O\left(t,t_0\right)\,,
\ee
and
\be
O\left(t_0-t_0\right)=1\,,
\ee
being simply a propagator, such that the solution can be cast as
\be
x(t)=O\left(t-t_0\right) x\left(t_0\right)\,.
\ee
Actually, $O$ is a matrix whose columns are the two linear independent solutions, $x_1$ and $x_2$ satisfaying
\be
x_1(t_0 )=\left(\begin{array}{c}1\\0\end{array}\right)\,,~~\text{and}~~x_2(t_0 )=\left(\begin{array}{c}0\\1\end{array}\right)\,,
\ee
and therefore at $t_0$ the Wroskian is
\be
det O\left(t_0,t_0\right)=1.
\ee
Abel's identity relates this to the late time Wroskian. In our case, where we neglect friction terms encoded in the Hubble expansion, this leads to
\be\label{eq:Wroskiancond}
detO\left(t,t_0\right)=1\,.
\ee
Floquet's theorem states that for periodic $E(t+T)=E(t)$
\be
O\left(t,t_0\right)=P\left(t,t_0\right)e^{\left(t-t_0\right) \mathcal{M}\left(t_0\right)}\,,
\ee
with periodic $P\left(t+T,t_0\right)=P\left(t,t_0\right)$ and $\mathcal{M}\left(t_0\right)$ such that $O\left(\left(t_0+T\right),t_0\right)=\exp \left(T\mathcal{M}\left(t_0\right)\right)$. The eigenvalues $\mu _1$ and $\mu _2$ of $\mathcal{M}(t_0)$ are called Floquet exponents, that from $det O\left(t,t_0\right)=1$, satisfy $$\mu_1+\mu_2=0\,,$$ 
and therefore there is only one linear independent that we denote by $\mu$. With $e_{\pm }$ the corresponding eigenvectors, neglecting the possibility $\mu=0$, then the solutions are
\be
x(t) = c_+\mathcal{P}_+(t, t_0)e^{\mu (t-t_0)}+c_-\mathcal{P}_-(t,
t_0)e^{-\mu (t-t_0)}\,,
\ee
where $\mathcal{P}_{\pm }=P\left(t,t_0\right)e_{\pm }\left(t_0\right)$. This can be also expresed in terms of the $O\left(t+T,t_0\right)$ eigenvalues, called Floquet multipliers, and denoted by $\pi_\pm$, through
\be
\pi _{\pm }=e^{\pm T\mu}\,,
\ee
or
\be
\mu =\frac{1}{T}\left(\ln\left(\left|\pi _+\right|\right)+\rmi  \text{Arg}(\pi _+)\right)=\frac{1}{T}\left(\ln\left(\left|\pi _+\right|\right)-\rmi  \text{Arg}(\pi _.)\right)\,.
\ee
An exponential grow appear whenever $\Re(\mu)\neq0$ or $\left|\pi _\pm\right|\neq0$.

\subsection{Evaluating the exponents}
For a given problem follow the following steps:
\begin{enumerate}
	\item Solving the periodicity condition implies solving the e.o.m. for the homogeneous field neglecting the expansion of the universe, i.e. solve simultaneously
	\be
	\ddot{\phi}+V'(\phi)=0\,,\quad \dot a(t)=\sqrt{\frac{1}{3}\left(\frac12 \dot{\phi}^2+V(\phi)\right)}a(t)\,.
	\ee
	\item Determine the period for the potential
	\be
	T(\phi _{\max })=2	
	\int _{\phi _{\min }}^{\phi _{\max }}\frac{d \phi }{\sqrt{2 V\left( \phi _{\max }\right)-2 V\left( \phi \right)}}\,.
	\ee
	This depends on the initial amplitude $\phi_{max}=\phi\left(t_0\right)$ (assuming $\partial_t\phi(t_0)=0$), where the minimum $V\left(\phi_{max}\right)=V\left(\phi_{min}\right)$.
	\item Solve the e.o.m. for the fluctuations \eqref{eq:dletaphieom} from $t=t_0$, that we choose $t_0=0$, to $t=t_0+T$, for the two initial conditions $\{\delta \phi_{k,1}(t_0)=1\,,\partial_t\delta \phi_{k,1}(t_0)=0\}$ and $\{\delta \phi_{k,2}(t_0)=0\,,\partial_t\delta \phi_{k,2}(t_0)=1\}$. 
	\item Compute the Floquet exponents from the Floquet multipliers
	\be
	\pi^{\pm}_k=\frac{1}{2}\left.\left(\delta \phi_{k,1}+\delta \dot\phi_{k,2}\pm\sqrt {\left[\delta \phi_{k,1}-\delta \dot\phi_{k,2}\right]^2+4 \delta \phi_{k,2}\delta \dot\phi_{k,1}}\right)\right|_{t=t_0+T}\,.
	\ee
\end{enumerate}

\subsection{LVS and uplift}\label{app:LVSOsciup}
For this analysis we use the potential as appear in \eqref{eq:Vcritpoint}, with the canonical normalized field $\phi=\sqrt{\frac23}\vp$, changing the parameter $\gamma$ and performing for each case the analysis just explained. We chose $\tilde \alpha=1$ and the value of the position of the scalar potential minimum, though irrelevant for the analysis, is taken to be $\phi_*=5$. For the initial position we choose the midpoint between the inflation point and the barrier top position, i.e., $\phi_{in}=\phi_*+(\phi_{top}-\phi_{inf})/2$ . Given the decrease in the scalar potential convexity an increase in the Floquet exponents is expected for $\gamma\to 3$.  
As already reported in \cite{Antusch:2017flz} the values for the Floquet exponents are low enough to suspect the absence of oscillons, something also checked there by explicit numerical simulation. Figure \ref{fig:floquetLVSuplift} shows the results, where we see that the difference due to the uplifts is not relevant affecting the value only at a few percent, even for a large value like $\gamma=14/5$. 
\begin{figure}[htb]
	\begin{center}
		\includegraphics[width=0.8\textwidth
		]{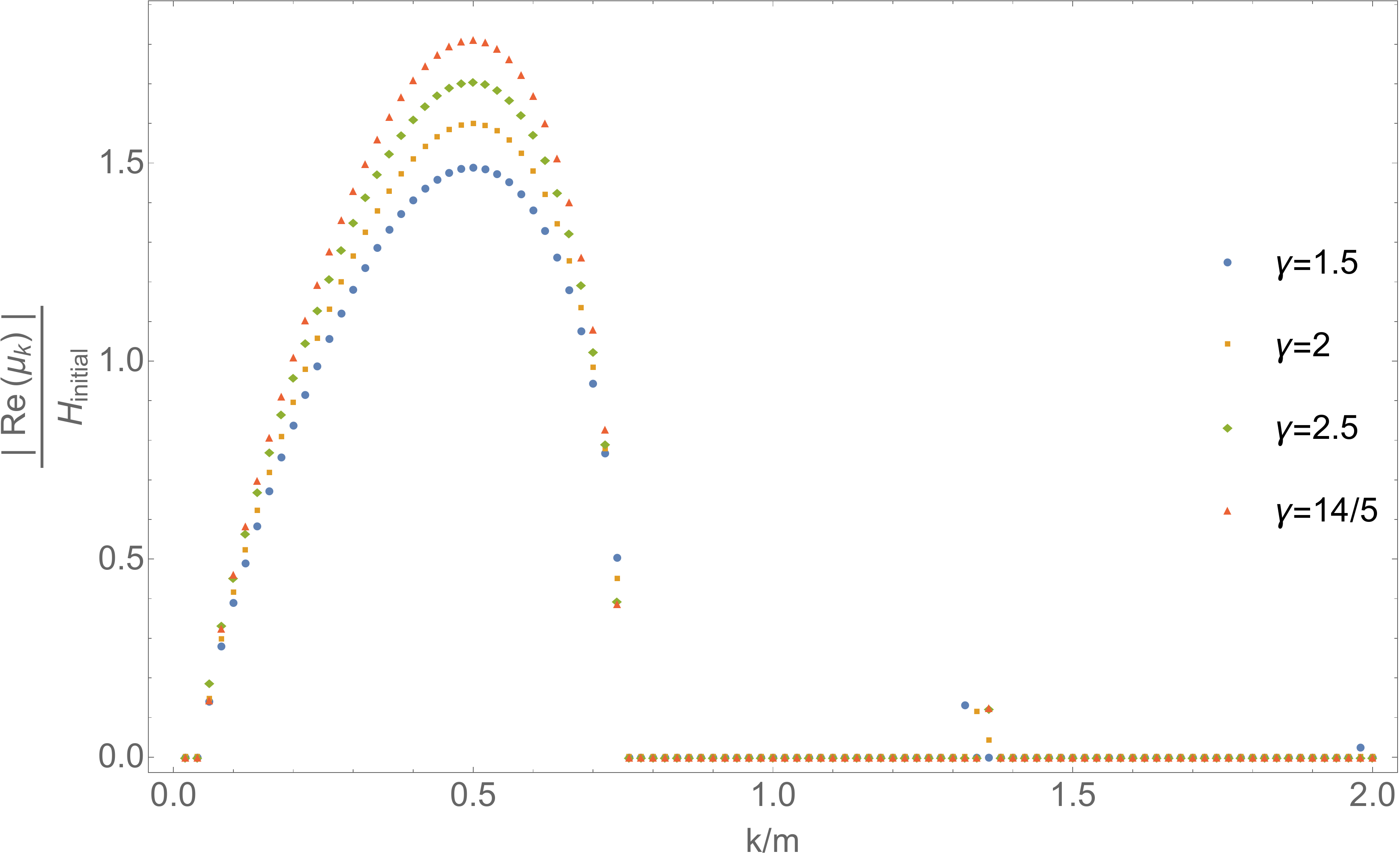}
	\end{center}
	\caption{Normalized real part of the Floquet exponents for the LVS scalar potential at zero temperature for different values of parameter $\gamma$ for an initial value of the amplitude given by $\phi_{in}=\phi_*+(\phi_{top}-\phi_{inf})/2$, the subscripts denoting the barrier top position and the inflection point. Even for the extreme value of $\gamma=14/5$ the size of the exponents is small enough to suspect the absence of oscillons for this potential. The dependency of the results on $\gamma$ is not dramatic though, as expected, the closer $\gamma$ to $3$ the larger the values are. }
\end{figure}\label{fig:floquetLVSuplift}

\bibliographystyle{JHEP}
\bibliography{refs}
\end{document}